\documentclass[apj]{emulateapj}

\usepackage{bm}
\usepackage{graphicx}
\usepackage{epsf}
\usepackage{graphics}
\usepackage{amsmath}
\usepackage{amssymb}
\usepackage{multirow}
\usepackage{booktabs}
\usepackage{threeparttable}
\usepackage{float}
\usepackage{mathrsfs} 
\DeclareMathSizes{15}{1}{1}{1}

\newcommand{\msunh}{\>h^{-1}\rm M_\odot}

\newcommand{\Msun}{\>{\rm M_{\odot}}}

\newcommand{\mpch}{\>h^{-1}{\rm {Mpc}}}

\newcommand{\kmsmpc}{\>{\rm km}\,{\rm s}^{-1}\,{\rm Mpc}^{-1}}

\newcommand{\rmag}{\>^{0.1}{\rm M}_r-5\log h}

\newcommand{\rmnp}{{\rm p}}
\newcommand{\rmm}{{\rm m}}
\newcommand{\rmh}{{\rm h}}

\def\gtsima{$\; \buildrel > \over \sim \;$}
\def\ltsima{$\; \buildrel < \over \sim \;$}
\def\prosima{$\; \buildrel \propto \over \sim \;$}
\def\gsim{\lower.7ex\hbox{\gtsima}}
\def\lsim{\lower.7ex\hbox{\ltsima}}
\def\simgt{\lower.7ex\hbox{\gtsima}}
\def\simlt{\lower.7ex\hbox{\ltsima}}
\def\simpr{\lower.7ex\hbox{\prosima}}
\def\la{\lsim}
\def\ga{\gsim}
\def\lta{\la}
\def\gta{\ga}

\usepackage{color}

\shorttitle{Mapping the real space distributions of galaxies}
\shortauthors{Shi et al.}

\begin{document}

\title{  Mapping the Real Space Distributions of Galaxies in SDSS DR7: I. Two Point
Correlation Functions }

\author{Feng Shi\altaffilmark{1,8}, Xiaohu Yang\altaffilmark{2,3},
  Huiyuan Wang\altaffilmark{4}, Youcai Zhang\altaffilmark{1},
  H.J. Mo\altaffilmark{5,6}, Frank C. van den Bosch\altaffilmark{7},
  Shijie Li\altaffilmark{1}, Chengze Liu\altaffilmark{2}, Yi
  Lu\altaffilmark{1}, Dylan Tweed\altaffilmark{2}, Lei
  Yang\altaffilmark{2} }

\altaffiltext{1}{Shanghai Astronomical Observatory, Nandan Road 80,
  Shanghai 200030, China; E-mail: sfeng@shao.ac.cn}

\altaffiltext{2}{Center for Astronomy and Astrophysics, Shanghai Jiao
  Tong University, Shanghai 200240, China;  E-mail: xyang@sjtu.edu.cn}

\altaffiltext{3}{IFSA Collaborative Innovation Center, Shanghai Jiao
  Tong University, Shanghai 200240, China}

\altaffiltext{4}{Key Laboratory for Research in Galaxies and Cosmology,
University of Science and Technology of China, Hefei, Anhui 230026, China}

\altaffiltext{5}{Department of Astronomy, University of Massachusetts,
Amherst MA 01003-9305, USA}

\altaffiltext{6}{Physics Department and Center for Astrophysics, Tsinghua University,
Beijing 10084, China}

\altaffiltext{7}{Department of Astronomy, Yale University, P.O. Box 208101,
  New Haven, CT 06520-8101, USA}

\altaffiltext{8}{University of Chinese Academy of Sciences, 19A,
  Yuquan Road, Beijing, China}

\begin{abstract}
  Using a method to correct redshift space distortion (RSD) for
  individual galaxies,  we mapped the real space distributions of
    galaxies in the Sloan Digital Sky Survey (SDSS) Data Release 7
    (DR7).  We use an ensemble of mock catalogs to demonstrate the
  reliability of our method.  Here as the first paper in a series,
    we mainly focus on the two point correlation function (2PCF) of
    galaxies.  Overall the 2PCF measured in the reconstructed real
    space for galaxies brighter than $\rmag=-19.0$ agrees with the
    direct measurement to an accuracy better than the measurement
    error due to cosmic variance, if the reconstruction uses the
  correct cosmology.  Applying the method to the SDSS DR7, we
  construct a real space version of the main galaxy catalog, which
  contains 396,068 galaxies in the North Galactic Cap with redshifts
  in the range $0.01 \leq z \leq 0.12$.  The Sloan Great Wall, the
  largest known structure in the nearby Universe, is not as dominant
  an over-dense structure as appears to be in redshift space. We
  measure the 2PCFs in reconstructed real space for galaxies of
  different luminosities and colors. All of them show clear deviations
  from single power-law forms, and reveal clear transitions from
  1-halo to 2-halo terms.  A comparison with the corresponding 2PCFs
  in redshift space nicely demonstrates how RSDs boost the clustering
  power on large scales (by about $40-50\%$ at scales $\sim10 \mpch$)
  and suppress it on small scales (by about $70-80\%$ at a scale of
  $0.3 \mpch$). 
\end{abstract}

\keywords {methods: statistical - galaxies: haloes - dark matter -
  large-scale structure of Universe }

\section{Introduction}
\label{sec:intro}

One of the important properties of the galaxy population is the
distribution of galaxies in space \citep[e..g.][]{peebles80, MBW10}.
This distribution can be used to infer the large scale mass
distribution in the universe, thereby constraining cosmological models
\citep[e.g.][]{Fisher1994, Pea2001, Hawkins2003, Yang2004,
  Tinker2005}. Furthermore, the spatial clustering of galaxies is also
one of the key pieces of observational data to establish the relation
between galaxies and dark matter (halos) statistically
\citep[e.g.][]{Jing1998, Pea2000,Yang2003,Yang2012}, and to understand
how galaxies form and evolve in the cosmic density field.

One of the main goals of large redshift surveys of galaxies, such as
the 2 degree Field Galaxy Redshift Survey \citep[2dFGRS;][]{Col2001}
and the Sloan Digital Sky Survey \citep[SDSS;][]{York2000} is,
therefore, to provide a data base to study the three dimensional
distribution of galaxies as accurately possible. However, a key
problem this endeavor is that redshifts of galaxies are not exact
measures of distances due to the peculiar motions of galaxies.  The
spatial distribution and clustering of galaxies observed in redshift
space are thus distorted with respect to the real-space distribution
and clustering \citep[e.g.][]{Sar1977,Dav1983, Kai1987, Reg1991,
  Ham1992, Wey1993}.  Take the two-point correlation function (2PCF)
of galaxies as an example.  The 2PCF in the 2-dimensional space, with
1 dimension along the line-of-sight and the other in the perpendicular
direction, appears elongated on small scales and squashed on large
scales along the line-of-sight direction, in contrast to an isotropic
pattern expected from a statistically homogeneous and isotropic
distribution in real space.  Such anisotropies are clearly produced by
redshift distortions and need to be corrected in order to get the true
distribution of galaxies in space.  Theoretically, models of the
pairwise peculiar velocities of galaxies have been used to model the
effects of redshift distortions on the measured 2PCF in redshift space
\citep[e.g.][]{Dav1983, Fisher1994, Jing1998}.  Alternatively, one
simply measures the projected 2PCF and uses it to infer the
three-dimensional 2PCF \citep[e.g.][]{Jing1998, Li2006, Zehavi2011}.

In the gravitational instability scenario of structure formation, the
redshift distortion is not just a contamination one has to correct in
order to get the real clustering of galaxies, it in fact contains
useful information about cosmology as well as the mass distribution in
the universe. On large scales, the infall motions of galaxies, which
produce the squashing in the 2D redshift-space 2PCF \citep[the Kaiser
effect,][]{Kai1987}, is linearly proportional to the amplitudes of the
mass density fluctuations on large-scales.  In this case, one can
compute the quadrupole-to-monopole ratio of the 2D 2PCF to get $\beta
\equiv f(\Omega_\rmm)/b$, where $\Omega_\rmm$ is the density parameter
of mass, and $b$ is the effective linear bias of the galaxies in
question \citep[e.g.][and references therein]{Guz2008, Sam2012,
  Dawson2016}.  When the measurement is combined with weak
gravitational lensing results, it can also be used as a sensitive
probe of (modified) gravitation theories on cosmology scales
\citep{Zhang07, Reyes10, Blake16}. On smaller scales, the modeling of
redshift space distortion  \citep[the Finger of God
  effect,][]{Jac1972, Tul1978} is complicated by the nonlinear
mapping between real-space and redshift-space. Great efforts have been
made not only to understand its impacts on galaxy clustering
\citep[e.g.][]{Zhang13,Zheng13,Zhang15,Zheng15a,Zheng15b}, but also to
extract useful cosmological information (Mo, Jing \& Boerner 1993;
Jing, Mo \& Boerner 1998; Yang et al. 2004; Li et al. 2012).

The approaches adopted earlier to deal with redshift distortions in
galaxy clustering have been hampered by the fact that the large-scale
Kaiser effect and the small-scale Finger of God effect are interwoven,
and models based on a simple pairwise peculiar velocity distribution
can only be served as an approximation.  The situation is complicated
even more by the fact that the effect bias in galaxy distribution may
be nonlinear and its form is not known {\it a priori}. Models based on
the projected correlation function has its own problem, because the
projection mixes clustering on different scales so that the conversion
from the projected function to the three dimensional function can be
uncertain.  Thus, in order to make full use of galaxy redshift surveys
to study the large-scale structure of the universe, a change of
tactics is needed.

One possible way is first to make corrections of redshift distortions
for individual galaxies, and then use the `pseudo' real space
distribution of galaxies to derive statistical measures of galaxy
clustering in real space. As mentioned above, redshift distortions are
of two different kinds. One is the Kaiser effect produced by the
coherent flow due to the gravitational action of large-scale structure
\citep[][]{Kai1987}, the other is the Finger of God (FOG) effect
generated by the random motions of galaxies within virialized halos on
small scales. To deal with the FOG effect, \citet{Teg2002} used an
friends-of-friends method to link galaxies and suppressed the
over-density of the pairs along the line of sight by a factor of 10.
They applied this FOG suppression to the 2dFGRS \citep{Teg2002} and
SDSS \citep{Tegmark2004} in their estimates of the power spectra of
galaxy distribution.  In a paper aimed at reconstructing the cosmic
web from 2dFGRS, \citet{Erdogdu2004} attempted to dealt with the FOG
effect by compressing 25 fingers seen in redshift space using groups
identified by \citet{Eke2004}.  For the Kaiser effect,
\citet{Yahil1991} used a bias model to get the density field from the
galaxy distribution and iteratively corrected the infall motions of
galaxies.  Along the same line, a number of approaches have been taken
to recover/correct the infall motions on the basis of galaxy
distribution \citep[e.g.][]{Monaco1999, Lavaux2008, WH2009,
  Branchini2012, WH2012, Kitaura2012, Granett2015, Jasche2015,
  Kitaura2016, Ata2016}. In particular, \citet{WH2009,WH2012} used
galaxy groups as proxies of dark matter halos to reconstruct the
density field, which in turn was used to obtain the velocity field.

So far there has been no real attempt to correct for both the large
scale velocities and small scale random motions of galaxies in a
systematic way.  The main purpose of the present paper is to carry out
such an investigation, using galaxies observed in the SDSS DR7, which
is still among the best redshift surveys available.  Based on this
galaxy catalog, \citet[hereafter Y07]{Yang2007} have constructed a
galaxy group catalog using an adaptive halo-based group finder
\citep[see also][]{Yang2005}. Detailed tests with mock galaxy
catalogues have shown that the group finder is very successful in
associating galaxies according to their common dark matter halos.  In
particular, the group finder performs reliably not only for rich
systems, but also for poor systems, including isolated central
galaxies in low mass halos. The reliable memberships of galaxies in
groups provide a unique opportunity to correct for the FOG effects for
individual galaxy systems. In addition, as shown in \citet[hereafter
W12]{WH2012}, the group catalog can also be used to reconstruct the
mass density, tidal and velocity (MTV) fields in the SDSS DR7 volume,
using the halo-domain method developed in \citet{WH2009}. Since the
relation between halo and mass distributions is better understood than
that between galaxies and mass, the mass and velocity fields
constructed are much more accurate than those constructed directly
from the galaxy distribution. The redshift distortions on large scales
can, therefore, also be modeled accurately for individual galaxies.
With all these, we can obtain a catalog of galaxies in quasi-real
space.  We can then not only examine in detail various types of
redshift distortions, but also measure the real space clustering of
galaxies.

This paper is organized as follow. In Section 2 we present the galaxy
and group catalogs used in this paper. Section 3 introduces the
methods to correct for the redshift distortions and to characterize
the galaxy clustering. In Section 4 we use mock galaxy catalogs to
test the reliability of our correction model. The application to the
SDSS DR7 and the results are presented in Section 5. Finally, we
summarize our main findings in Section 6. Throughout this paper,
unless stated otherwise, physical quantities are quoted using the
WMAP9 cosmological parameters \citep{Hin2013}: $\Omega_\rmm = 0.282$,
$\Omega_{\Lambda} = 0.718$, $\Omega_{\rm b} = 0.046$, $n_{\rm
  s}=0.965$, $h=H_0/(100 \kmsmpc) = 0.697$ and $\sigma_8 = 0.817$.

\begin{table*}
\center
\scalebox{0.9}{
\begin{threeparttable}[c]
\caption{Galaxy Subsamples }\label{subsamp}
\setlength{\tabcolsep}{3pt}
\begin{tabular}{ccccccc}
\toprule
\multirow{2}{*}{Absolute Magnitude} &
\multicolumn{3}{c}{\multirow{2}{*}{Flux-limited}} & 
\multicolumn{3}{c}{\multirow{2}{*}{Volume-limited}} \\\\ 

\multirow{2}{*}{$\rmag$ }  &
\multirow{2}{*}{Redshift}  & 
\multirow{2}{*}{~~~$N_{gal}$($N_{blue}$/$N_{red}$)}  &
\multirow{2}{*}{Averaged Magnitude} &
\multirow{2}{*}{Redshift}  & 
\multirow{2}{*}{~~~$N_{gal}$($N_{blue}$/$N_{red}$)} &
\multirow{2}{*}{Averaged Magnitude} \\\\

\hline
\multirow{2}{*}{$\bm{[}-23,-22\bm{]}$} & 
\multirow{2}{*}{$\bm{[}0.01, 0.12\bm{]}$}  & 
\multirow{2}{*}{$~~~2200(379/1821)$} &
\multirow{2}{*}{$-22.22$} &
\multirow{2}{*}{$\bm{[}0.01, 0.12\bm{]}$}  & 
\multirow{2}{*}{$~~~2200(379/1821)$} &
\multirow{2}{*}{$-22.22$} \\

\multirow{2}{*}{$\bm{[}-22,-21\bm{]}$} & 
\multirow{2}{*}{$\bm{[}0.01, 0.12\bm{]}$}  & 
\multirow{2}{*}{$~~~42207(11997/30210)$} &
\multirow{2}{*}{$-21.34$} &
\multirow{2}{*}{$\bm{[}0.01, 0.12\bm{]}$}  & 
\multirow{2}{*}{$~~~42207(11997/30210)$} &
\multirow{2}{*}{$-21.34$} \\

\multirow{2}{*}{$\bm{[}-21,-20\bm{]}$} & 
\multirow{2}{*}{$\bm{[}0.01, 0.12\bm{]}$}  & 
\multirow{2}{*}{$~~~156765(64956/91809)$}  &
\multirow{2}{*}{$-20.44$}  &
\multirow{2}{*}{$\bm{[}0.01, 0.113\bm{]}$}  & 
\multirow{2}{*}{$~~~134801(55572/79229)$} &
\multirow{2}{*}{$-20.43$}  \\
                                       
\multirow{2}{*}{$\bm{[}-20,-19\bm{]}$} & 
\multirow{2}{*}{$\bm{[}0.01, 0.12\bm{]}$}  & 
\multirow{2}{*}{$~~~127444(71018/56426)$} &
\multirow{2}{*}{$-19.57$}  &
\multirow{2}{*}{$\bm{[}0.01, 0.075\bm{]}$}  & 
\multirow{2}{*}{$~~~73391(41659/31732)$} &
\multirow{2}{*}{$-19.47$}  \\
 
\multirow{2}{*}{$\bm{[}-19,-18\bm{]}$} & 
\multirow{2}{*}{$\bm{[}0.01, 0.12\bm{]}$}  & 
\multirow{2}{*}{$~~~43894(31646/12248)$}  &
\multirow{2}{*}{$-18.58$}  &
\multirow{2}{*}{$\bm{[}0.01, 0.045\bm{]}$}  & 
\multirow{2}{*}{$~~~21875(16052/5823)$} &
\multirow{2}{*}{$-18.48$}  \\

\multirow{2}{*}{$\bm{[}-18,-17\bm{]}$} & 
\multirow{2}{*}{$\bm{[}0.01, 0.12\bm{]}$}  & 
\multirow{2}{*}{$~~~17259(14327/2932)$}  &
\multirow{2}{*}{$-17.57$}  &
\multirow{2}{*}{$\bm{[}0.01, 0.026\bm{]}$}  & 
\multirow{2}{*}{$~~~5618(4818/800)$} &	
\multirow{2}{*}{$-17.46$}  \\
\\ \bottomrule
\end{tabular}
\end{threeparttable}}
\end{table*}

\section{The SDSS galaxy and group catalogs}
\label{sec:data}

The galaxy sample used in this paper is constructed from the New York
University Value-Added Galaxy Catalog \citep[NYU-VAGC;][]{Bla2005},
which is based on SDSS DR7 \citep{Aba2009}, but with an independent
set of significantly improved reductions over the original
pipeline. In addition, as galaxy groups play a key role in our
approach to correct for the redshift distortions, we make use of the
group catalog constructed in \citep[hereafter Y12]{Yang2012} for SDSS
DR7.  This group catalog is based on all galaxies in the Main Galaxy
Sample with extinction-corrected apparent magnitude brighter than $r =
17.72$, with redshifts in the range $0.01 \leq z \leq 0.20$ and with a
redshift completeness $ C_z > 0.7$. The catalog contains a total of
639,359 galaxies with a sky coverage of 7,748 deg$^2$.  Moreover, the
galaxy catalog mainly covers two sky regions: a larger contiguous
region in the Northern Galactic Cap (NGC) and a smaller three-stripe
region in the Southern Galactic Cap (SGC).  The former contains
584,473 galaxies with a sky coverage of 7,047 deg$^2$.

Based on this SDSS DR7 galaxy catalog, Y12 used the adaptive
halo-based group finder developed by Y05 to select galaxy groups. This
group finder has been applied to the SDSS DR4 in Y07.  Following Y07,
the masses of the associated dark matter halos are estimated based on
the ranking of the total characteristic luminosities of groups or the
total characteristic stellar masses using group member galaxies more
luminous than $\rmag=-19.5$.  Both halo masses agree very well with
each other, and we adopt the halo masses based on the characteristic
luminosity ranking in this paper.  In addition, we have updated group
membership as well as halo masses according to WMAP9 cosmology.

Using this group catalog, W12 reconstructed the velocity field, which
we use in this paper to correct for the redshift space distortions.
The method of W12 explicitly depends on the density field as
represented by dark matter halos above a given mass threshold, $M_{\rm
  th}$.  We adopt $M_{\rm th}=10^{12.5}h^{-1}M_\sun$ and so, to be
complete, restrict our sample to the nearby volume covering the
redshift range $0.01 \leq z \leq 0.12$\footnote{In practice, to keep
  large scale mode at the $z=0.12$, we use groups in the redshift
  range $0.01 \leq z \leq 0.13$ for our velocity reconstruction.  }.
In addition, since the W12 reconstruction method can be significantly
impacted by survey boundaries, we focus only on the more contiguous
NGC region.

Applying all these selection criteria to the galaxy and group catalogs
leaves us with a set of 286,043 groups, hosting a total of 396,068
galaxies in the NGC region with redshifts in the range $0.01\leq z
\leq 0.12$. Finally, using this sample we construct both flux-limited
and volume-limited subsamples for galaxies in the following six
absolute $r$-band magnitude bins: $\rmag=$ $\bm{[}-23.0,-22.0\bm{]}$,
$\bm{[}-22.0,-21.0\bm{]}$, $\bm{[}-21.0,-20.0\bm{]}$,
$\bm{[}-20.0,-19.0\bm{]}$, $\bm{[}-19.0,-18.0\bm{]}$ and
$\bm{[}-18.0,-17.0\bm{]}$.  The corresponding redshift ranges, numbers
of galaxies and averaged magnitude are indicated in
Table~\ref{subsamp}.  These luminosity samples are further divided
into blue and red subsamples, as detailed in Section
\ref{sec:clusan}. Note that there is no difference in the redshift
limit between the flux-limited and volume-limited for the first two
brightest samples, because all the galaxies with such luminosities can
be observed to $z=0.12$.  For a fainter sample, even the brightest
galaxies in the luminosity bin can be observed only to redshift
$z<0.12$.  In most cases we only show results obtained from the flux
limited samples, because the results obtained from the volume limited
samples are very similar. Note also that in the reconstructed real
space, which we will perform later, the number of galaxies in a sample
will change very slightly.

\begin{table*}
\scalebox{1.2}{
\begin{threeparttable}
\caption{Description of different spaces.}\label{spaces}
\centering

\setlength{\tabcolsep}{20pt}
\begin{tabular}{ll}
\toprule
SPACE     &  DESCRIPTION \\
\hline
\multirow{2}{*}{Real space} & 
\multirow{2}{*}{survey geometry without redshift distortions}  \\
\\ \hline
\multirow{2}{*}{FOG space} & 
\multirow{2}{*}{distorted only by FOG effect:
 ~~$z_{\rm obs}=z_{\rm cos}+\frac{v_{\sigma}}{c}(1+z_{\rm cos})$ } \\
\\ \hline
\multirow{2}{*}{Kaiser space} & 
\multirow{2}{*}{distorted only by Kaiser effect:
 ~~$z_{\rm obs}=z_{\rm cos}+\frac{v_{\rm cen}}{c}(1+z_{\rm cos})$} \\
\\ \hline
\multirow{2}{*}{Redshift space} & 
\multirow{2}{*}{distorted by both Kaiser and FOG effects:
 ~~$z_{\rm obs}=z_{\rm cos}+\frac{v_{\rm pec}}{c}(1+z_{\rm cos})$ } \\
\\ \toprule
\multirow{2}{*}{Re-real space} & 
\multirow{2}{*}{reconstructed real space; based on correcting redshift space distortions } \\
\\ \hline
\multirow{2}{*}{Re-Kaiser space} & 
\multirow{2}{*}{reconstructed Kaiser space; based on correcting for FOG effect only } \\
\\ \hline
\multirow{2}{*}{Re-FOG space} & 
\multirow{2}{*}{reconstructed FOG space; based on correcting for Kaiser effect only } \\
\\ \bottomrule
\end{tabular}
\textbf{Notes.} The first four spaces are `true' spaced, based on true
groups (all galaxies belonging to the same dark matter halo). The
final three space are `reconstructed' spaces based on groups
identified applying the group finder in redshift space.
\end{threeparttable}}
\end{table*}

\section{METHODOLOGY AND BASIC ANALYSIS}
\label{sec:method}

We now turn to our main goal: correcting the SDSS redshifts for
redshift space distortions induced by peculiar velocities, thus
allowing for a direct measurement of the two-point correlation
functions of galaxies in real space.  Before delving into details, we
first introduce some concepts regarding redshift space distortions and
our approach to correct for them.

\subsection{Redshift Space Distortions}
\label{sec:rsd}

In the absence of peculiar velocities, the redshift of a galaxy, $z$,
is directly related to its comoving distance, $r$. For a flat
Universe, this relation is given by
\begin{equation}
\label{distance_z}
r(z)=\frac{1}{H_0}\int_0^z
\frac{{\rm d}z}{\sqrt[]{\Omega_\Lambda+\Omega_\rmm(1+z)^3}},
\end{equation}
with $H_0$ the Hubble constant. In reality, though, the observed
redshift of a galaxy, $z_{\rm obs}$, consists of a cosmological
contribution, $z_{\rm cos}$, arising from the Hubble expansion plus
a Doppler contribution, $z_{\rm pec}$, due to the galaxy's peculiar
velocity along the line-of-sight, $v_{\rm pec}$. In the non-relativistic
case we have that
\begin{equation}\label{zobs}
  z_{\rm obs} = z_{\rm cos} + z_{\rm pec} = z_{\rm cos} + \frac{v_{\rm pec}}{c}(1+z_{\rm cos}),     
\end{equation}
with $c$ the speed of light.

The redshift distance, $r(z_{\rm cos})$, of a galaxy inferred from its
observed redshift differs from its true comoving distance, which is
given by $r(z_{\rm cos})$. Hence, peculiar velocities give rise to
redshift space distortions (RSDs), which complicate the interpretation
of galaxies clustering but also contain important additional
information about the cosmic mass distribution. After all, the
peculiar velocities are induced by this matter distribution, which is
itself correlated with the distribution of galaxies.  On small scales
the virialized motion of galaxies within dark matter halos cause a
reduction of the correlation power, known as the finger-of-God (FOG)
effect, while on larger scales the correlations are boosted due to the
infall motion of galaxies towards overdensity regions, know as the
Kaiser effect \citep{Kai1987}.

Since each galaxy is believed to reside in a dark matter halo, it is
useful to split the peculiar velocity of a galaxy into two components:
\begin{equation}
v_{\rm pec} = v_{\rm cen} + v_{\sigma}\,.
\end{equation} 
Here $v_{\rm cen}$ is the line-of-sight velocity of the center of the
halo, and $v_{\sigma}$ is the line-of-sight component of the velocity
vector of the galaxy with respect to that halo center. Roughly
speaking, $v_{\rm cen}$ is a manifestation of the Kaiser effect (at
least on large scales), while $v_{\sigma}$ mainly contributes to the
FOG effect.  Hence, for convenience in what follows, we define the
Kaiser and FOG redshifts as
\begin{equation}\label{zkaiser}
z_{\rm Kaiser} = z_{\rm cos}+\frac{v_{\rm cen}}{c}(1+z_{\rm cos}),     
\end{equation} 
\begin{equation}\label{zfog}
z_{\rm FOG} = z_{\rm cos}+\frac{v_{\sigma}}{c}(1+z_{\rm cos}).
\end{equation}

The various redshifts thus defined, allow us to define a number of
different spaces, in addition to the standard real and redshift
spaces.  Table~\ref{spaces} gives a brief description of the various
spaces used in this study. In each space, galaxy distances are
computed using their corresponding redshifts injected into Eq.~(1).
All spaces have the geometry of the SDSS DR7. The top four spaces
listed, are based on true velocities and true groups (dark matter
halos), without observational errors, or errors in group
identifications and/or membership.  The bottom three spaces (those
starting with `Re'), on the other hand, are reconstructed spaces,
obtained by correcting for the corresponding redshift
distortions. These are based on the reconstructed velocity field, and
on groups identified applying the group finder in redshift space (see
\S\ref{sec:correction} below). In what follows, we refer to the top
four spaces as `true' spaces, and the lower three spaces as
`reconstructed' spaces.

\subsection{Correcting for redshift space distortions}
\label{sec:correction}

We now describe our method to correct the redshifts in the SDSS DR7
survey volume for redshift space distortions. The method separately
treats the Kaiser effect and the FOG effect, as detailed below.

\subsubsection{Correcting for the Kaiser effect}
\label{sec:KaiserCORR}

In order to correct for the Kaiser effect, we reconstruct the velocity
field in the linear regime using the method of W12. Here we briefly
summarize the main ingredients of this reconstruction method, and
refer the reader to W12 for more details.  In the linear regime, the
peculiar velocities are induced by, and proportional to, the
perturbations in the matter distribution. If we write the velocity
field, $\bm{v}(\bm{x})$, as a sum of Fourier modes,
\begin{equation}
\bm{v}(\bm{x}) = \sum_{k} \bm{v}(k) \, e^{i \bm{k} \cdot \bm{x}},
\end{equation}
then, in the linear regime, each mode can be written as 
\begin{equation}\label{eq_vk}
\bm{v}(\bm{k}) = H \, a \, f(\Omega) \, \frac{i\bm{k}}{k^2} \, \delta(\bm{k}).
\end{equation}
Here $H = \dot{a}/a$ is the Hubble parameter, $a$ is the scale factor,
$\delta(\bm{k})$ is the Fourier transform of the density perturbation
field $\delta(\bm{x})$, and $f(\Omega) = d\ln D/d\ln a \simeq
\Omega^{0.6}_\rmm + \frac{1}{70} \Omega_\Lambda(1 + \Omega_\rmm/2)$
\citep[e.g.][]{Lah1991}.

Hence, for a given cosmology one can directly infer the linear
velocity field from the density perturbation field, $\delta(\bm{x})$.
The challenge, however, is to reconstruct the matter field from
observations in redshift space. The unique aspect of the W12 method is
that it doesn't try to reconstruct $\delta$, but instead focuses on
the matter density field, $\delta_\rmh$, which is the (large scale)
matter distribution due to dark matter halos with a mass $M_\rmh \geq
M_{\rm th}$. As is well known, dark matter halos are biased tracers of
the mass distribution \citep[e.g.,][]{MoWhite96}.  On large, linear
scales we have that $\delta_\rmh(\bm{x}) = b_{\rm hm} \delta(\bm{x})$,
where $b_{\rm hm}$ is the linear bias parameter for dark matter halos
with mass $M_\rmh \geq M_{\rm th}$, which is given by
\begin{equation}\label{bhm}
  b_{\rm hm} = {\int_{M_{\rm th}}^{\infty} M \, b_\rmh(M) \, n(M) \, {\rm d}M \over
\int_{M_{\rm th}}^{\infty} M \, n(M) \, {\rm d}M}  
\end{equation}
where $n(M)$ and $b_\rmh(M)$ are the halo mass function and the
halo bias function, respectively. Hence, one can reconstruct the
peculiar velocity field (on linear scales) from $\delta_{\rm
  h}(\bm{x})$ using
\begin{equation}\label{vrecover}
  \bm{v}(\bm{k}) = H \, a \, f(\Omega) \,
  \frac{i\bm{k}}{k^2} \, {\delta_\rmh(\bm{k}) \over b_{\rm hm}}\,.
\end{equation}
In other words, the velocity field can be reconstructed even if we
only have the distribution of dark matter halos above some mass
threshold. This is fortunate, since it means that we can use our
galaxy group catalog, in which galaxy groups are linked with dark
matter halos above some mass threshold.

In order to reconstruct the velocity field in the SDSS survey volume,
we proceed as follows. We first embed the survey volume in a period,
cubic box of $726 \mpch$ on a side. The size of this `survey box' is
chosen to be about $100 \mpch$ larger than the maximum scale of the
survey volume among the three axes. Next, we divide the box into
$1024^3$ grid cells, and use groups with an assigned mass $M_\rmh \geq
M_{\rm th} = 10^{12.5} \msunh$ to compute $\delta_\rmh(\bm{x})$ on
that grid using the method described in detail in W12. In order to
suppress non-linear velocities that are not captured by the linear
model, we smooth $\delta_\rmh(\bm{x})$ using a Gaussian smoothing
kernel with a mass scale of $10^{14.75} \msunh $
\citep[see][]{WH2009}. Next, we Fast Fourier Transform (FFT) this
smoothed overdensity field, and compute $\bm{v}(\bm{k})$ using
Eq.~(\ref{vrecover}), where $b_{\rm hm}$ is computed using
Eq.~(\ref{bhm}) adopting the halo mass and bias functions of
\citet{Tinker2008}.  Fourier transforming $\bm{v}(\bm{k})$ then yields
the velocity field, which we interpret as $\bm{v}_{\rm cen}(\bm{x})$,
the velocity field of group centers. Finally, the comoving distance of
each galaxy, corrected for the Kaiser effect, is computed as $r(z_{\rm
  corr})$ (cf. Eq.~[\ref{distance_z}]). Here
\begin{equation}\label{zcorr}
z_{\rm corr} = {z_{\rm obs} - (v_{\rm cen}/c) \over 1 + (v_{\rm cen}/c)}
\end{equation}
with $v_{\rm cen}$ the inferred line-of-sight velocity at the location
of the group to which the galaxy belongs. The location of the group
is defined as the luminosity weighted center of all group members.

Since the velocity field is computed using the redshift-space
distribution of the groups, this method needs to be iterated until
convergence is achieved. Using the inferred $\bm{v}_{\rm
  cen}(\bm{x})$, we correct the redshifts of all groups with an
inferred mass $M_\rmh \geq M_{\rm th}$ for their (inferred) peculiar
velocity, and recompute $\delta_\rmh(\bm{x})$ and $\bm{v}_{\rm
  cen}(\bm{x})$ using the same method. As shown in \cite{WH2009} and
\cite{WH2012}, typically two iterations suffice to reach convergence,
yielding an unbiased estimate of the linear velocity field.

\subsubsection{Correcting for the FOG effect}
\label{sec:FOGCORR}

The Finger-of-God effect arises due to the motion of galaxies inside
their dark matter halos. To first order, one can simply correct for
the FOG effect by assigning all group galaxies the redshift of the
group, and then computing the comoving distance using
Eq.~(\ref{distance_z}). However, this ignores the spatial extent of
dark matter halos, which can be quite substantial.

Unfortunately, it is impossible to infer a galaxy's line-of-sight
location from its peculiar velocity along that line-of-sight. Hence,
one can only correct for the FOG effect in a statistical sense, which
we do as follows. We assume that group galaxies are unbiased tracers
of the halo's mass distribution, and therefore follow a NFW
\citep{NFW1997}, radial number density profile
\begin{equation}\label{eq:NFW}
 n_{\rm gal}(r)=\frac{n_0}{(r/r_s)(1+r/r_s)^2}\,.
\end{equation}  
Here $r_{\rm s}$ is the characteristic radius and the normalization
parameter $n_0$ can be expressed in terms of the halo concentration
parameter $c=r_{180}/r_s$ as
\begin{equation}
  n_0 = {N_{\rm gal} \over 4 \pi \, r^3_{\rm s}} \, \left[
    \ln(1+c) - c/(1+c)\right]^{-1}
\end{equation}
Here $N_{\rm gal}$ is the number of group member galaxies, and
$r_{180}$ is the radius inside of which the halo has an average
overdensity of 180. Numerical simulations show that halo concentration
depends on halo mass, and we use the relation given by \citet{zhao09},
converted to the $c$ appropriate for our definition of halo mass.

In practice, we proceed as follows. We do not displace central
galaxies, which are defined to be the brightest group members.  For
satellite galaxies (all members other than centrals), we first
calculate the project distance $r_\rmnp$ between the galaxy and the
luminosity weighted center of its group. Then we randomly draw a
line-of-sight distance, $r_{\pi}$, for the galaxy whose probability
follows Eq.~\ref{eq:NFW} with $r=\sqrt{r_\rmnp^2+r_{\pi}^2}$. The
galaxy is then assigned a comoving distance given by $r(z_{\rm corr})
+ r_{\pi}$, with the $z_{\rm corr}$ of Eq.~(\ref{zcorr}). We have
verified that using the location of the central galaxy, rather than
the luminosity weighted center of the group, yields results that are
virtually indistinguishable.

\subsection{Two-point correlation functions}
\label{sec:2pf}

In this paper, we use 2PCFs to characterize the clustering of
galaxies.  We estimate the two-dimensional 2PCF, $\xi(r_{\rm
  p},r_\pi)$, for galaxies in each sample using the following
estimator:
\begin{equation}\label{eq:2pcf}
\xi(r_\rmnp,r_\pi) = \frac{\langle RR \rangle
\langle DD \rangle}{\langle DR \rangle^2} - 1\,,
\end{equation}
where $\langle DD \rangle$, $\langle RR \rangle$ and $\langle DR
\rangle$ are, respectively, the number of galaxy-galaxy, random-random
and galaxy-random pairs with separation $(r_\rmnp,r_\pi)$
\citep{Ham1993}.  The variables $r_\rmnp$ and $r_\pi$ are the pair
separations perpendicular and parallel to the line-of-sight,
respectively. Explicitly, for a pair of galaxies, one located at $s_1$
and the other at $s_2$,   where $s_i$ is computed using
  Eq.~(\ref{distance_z}) , we define
\begin{equation}
r_\pi =\frac{\bm{s}\cdot\bm{l}}{\bm{\mid l \mid}},
\quad r_\rmnp = \sqrt{\bm{s} \cdot \bm{s}-r_{\pi}^2}\,.
\end{equation}
Here $\bm{l} = (\bm{s_1}+\bm{s_2})/2$ is the line of sight
intersecting the pair and $\bm{s}=\bm{s_1}-\bm{s_2}$.

The projected 2PCF, $w_\rmnp(r_\rmnp)$ is estimated using
\begin{equation}\label{eq:wrp}
 w_\rmnp(r_\rmnp)=\int_{-\infty}^{\infty}\xi(r_\rmnp,r_\pi) \, {\rm d}r_\pi
 = 2 \sum \xi(r_\rmnp,r_\pi) \, \Delta r_\pi
\end{equation}
\citep{Dav1983}. In our analysis, the summation is over 100 bins of
$\Delta r_\pi = 1 \mpch$, corresponding to an integration from $r_\pi
= -100 \mpch$ to $+100 \mpch$.

The one-dimensional, redshift-space 2PCF, $\xi(s)$, is estimated by
averaging $\xi(r_\rmnp,r_{\pi})$ along constant $s=\sqrt{r^2_{\rm
    p}+r^2_{\pi}}$ using
\begin{equation}\label{eq:xir}
  \xi(s) = \frac{1}{2} \, \int_{-1}^{1}{\xi(r_\rmnp,r_{\pi}) \,
    {\rm d}\mu}\,,
\end{equation}
where $\mu$ is the cosine of the angle between the line-of-sight and
the redshift-space separation vector $\bm{s}$.  Alternatively, one can
also measure $\xi(s)$ by directly counting $\langle DD \rangle$,
$\langle RR \rangle$ and $\langle DR \rangle$ pairs as a function of
redshift-space separation $s$.

Whereas $\xi(r_\rmnp,r_\pi)$ and $\xi(s)$ are affected by RSDs, and
can therefore differ dramatically in different spaces (real space,
redshift space, Kaiser space, or FOG space), the projected correlation
function, which is integrated along the line-of-sight, is insensitive
to RSDs. In practice, though, since we only integrate over a finite
extent, the projected correlation function is hampered by residual
redshift space distortions (RRSDs). However, as we explicitly
demonstrate below, for an integration limit of $100 \mpch$ these RRSDs
are sufficiently small and do not significantly impact our results
\citep[see also][and references therein]{Bos2013}

\section{Tests based on mock data}
\label{sec:mock}
 
Before applying our reconstruction method to SDSS data, we test its
accuracy and reliability using a variety of mock SDSS DR7 surveys.  In
particular, we construct mock galaxy surveys in real space, Kaiser
space, FOG space and redshift space, which allows us to separately
test the corrections for the Kaiser and the FOG effects. In order to
gauge the accuracy of our reconstruction, we compare clustering
statistics from the reconstructed spaces with those obtained from
their respective true spaces.

Briefly, our tests therefore consist of the following four steps:
\begin{enumerate}
\item Construct mock galaxy samples in real, Kaiser, FOG and redshift
  space.
\item Run the galaxy group finder over each of these spaces.
\item Using these galaxy group catalogs, and the reconstruction
  methods described in \S\ref{sec:correction}, reconstruct the mock
  galaxy samples in re-Kaiser, re-FOG and re-real space by correcting
  for the Kaiser effect, the FOG compression, and both, respectively.
\item Measure the two-dimensional 2PCF $\xi(r_\rmnp,r_\pi)$, the
  projected 2PCF $w_\rmnp(r_\rmnp)$, and the redshift-space 2PCF
  $\xi(s)$, and compare the results from the reconstructed spaces with
  those from their corresponding true spaces.
\end{enumerate}

\subsection{The mock catalogs}
\label{sec:mockcat}

For our study, we use a high resolution $N$-body simulation which
evolves the distribution of $3072^{3}$ dark matter particles in a
periodic box of $500 \mpch$ on a side \citep{Li2016}. This simulation
was carried out at the Center for High Performance Computing at
Shanghai Jiao Tong University and was run with {\tt L-GADGET}, a
memory-optimized version of {\tt GADGET2} \citep{Spr2005}. The
cosmological parameters adopted by this simulation are consistent with
the WMAP9 results \citep{Hin2013}, and each particle has a mass of
$3.4 \times 10^{8}\msunh$. Dark matter halos are identified using the
standard friends-of-friends algorithm \citep[e.g.][]{Dav1985} with a
linking length that is 0.2 times the mean inter particle separation.
The mass of halos, $M_\rmh$, is simply defined as the sum of the
masses of all the particles in the halos, and we remove halos with
less than 20 particles. We refer to these halos as `real halos' in
what follows in order to distinguish them from the groups identified
by the group finder that is applied to the mock galaxy catalogs
described below.

Based on the halo catalog, we populate galaxies using the conditional
luminosity function (CLF) model of \citet{Yang2003}. The algorithm of
populating galaxies is similar to that outlined in \citet{Yang2004},
but here updated to the CLF in the SDSS $r$-band \citep[See][for a
  recent application]{Luyi2015}.  For completeness, we briefly
describe our method used to assign mock galaxies to our dark matter
halos.

We write the total CLF as the sum of a central galaxy and a satellite
galaxy component:
\begin{equation}\label{eq:CLF_fit}
\Phi(L|M_\rmh) = \Phi_{\rm cen}(L|M_\rmh) + \Phi_{\rm sat}(L|M_\rmh)\,.
\end{equation}
The central component is assumed to follow a log-normal distribution:
\begin{eqnarray}\label{eq:phi_c}
\Phi_{\rm cen}(L|M_\rmh) \, {\rm d}\log L
 ~~~~~~~~~~~~~~~~~~~~~~~~~~~~~~~~~~~~~~~~~~ && \\
=  {1 \over {\sqrt{2\pi}\sigma_{\rm c}}} {\rm exp}
\left[- { {(\log L  -\log L_{\rm c} )^2 } \over 2\sigma_{\rm c}^2} \right] \,
     {\rm d} \log L  \nonumber \,. &&
\end{eqnarray}
Here $\sigma_{\rm c}$ is a free parameter that expresses the scatter
in $\log L$ of central galaxies at fixed halo mass, and $\log L_{\rm
  c}$ is the expectation value for the (10-based) logarithm of the
luminosity of the central galaxy.  For the contribution from the
satellite galaxies we adopt a modified Schechter function:
\begin{eqnarray}\label{eq:phi_s}
\Phi_{\rm sat}(L|M) \,{\rm d} \log L
~~~~~~~~~~~~~~~~~~~~~~~~~~~~~~~~~~~~~~~~~~~~ && \\
= \phi^*_{\rm s}
\left ( {L\over L^*_{\rm s}}\right )^{(\alpha_{\rm s}+1)} {\rm exp} \left[- \left
  ({L\over L^*_{\rm s}}\right )^2 \right] \, \ln(10)  \, {\rm d} \log L
\nonumber \,. &&
\end{eqnarray}
Note that the parameters $L_{\rm c}$, $\sigma_{\rm c}$, $\phi^*_{\rm
  s}$, $\alpha_{\rm s}$ and $L^*_{\rm s}$ are all functions of the halo
mass $M_\rmh$.

Following \citet{Cac2009}, and motivated by the results of
\citet{Yang2008} and \citet{More2009}, we assume that $\sigma_{\rm c}$
is a constant (i.e., independent of halo mass), and that the $L_{\rm
  c}-M_\rmh$ relation has the following functional form,
\begin{equation}\label{eq:Lc_fit}
L_{\rm c} (M_\rmh) = L_0 \frac { (M_\rmh/M_1)^{\gamma_1} }{(1+M_\rmh/M_1)^{\gamma_1-\gamma_2} } \,.
\end{equation}
This model contains four free parameters: a normalized luminosity,
$L_0$, a characteristic halo mass, $M_1$, and two slopes, $\gamma_1$
and $\gamma_2$.  For satellite galaxies we use
\begin{equation}
\log L^*_{\rm s}(M_\rmh)  = \log L_{\rm c}(M_\rmh) - 0.25\,,
\end{equation}
\begin{equation}\label{alpha}
\alpha_{\rm s}(M_\rmh) = \alpha_{\rm s}
\end{equation}
(i.e., the faint-end slope of $\Phi_{\rm sat} (L|M_\rmh)$ is independent
of halo mass), and
\begin{equation}\label{phi}
\log[\phi^*_{\rm s}(M_\rmh)] = b_0 + b_1 (\log M_{12}) + b_2 (\log M_{12})^2\,,
\end{equation}
with $M_{12} = M_\rmh/(10^{12} h^{-1}\Msun)$. Thus defined, the CLF model
has a total of nine free parameters, characterized by the vector
\begin{equation}\label{lambdaCLF}
{\lambda}^{\rm CLF} \equiv (\log M_1, \log L_{0}, \gamma_1,
\gamma_2, \sigma_{\rm c}, \alpha_{\rm s}, b_0, b_1, b_2) \, .
\end{equation}
We emphasize that this functional form for the CLF accurately
describes the observational results obtained by \citet{Yang2008} from
the SDSS galaxy group catalog. The same functional form was adopted in
\citet{Cac2009} model galaxy-galaxy lensing, and, more recently, in
\citet{Bos2013}, \citet{More2013} and \citet{Cac2013} to
simultaneously constrain cosmological parameters and the galaxy-dark
matter connection using a combination of SDSS clustering and weak
lensing measurements.  Here we adopt the set of best-fit CLF
parameters listed in \citet{Cac2013} for cosmological parameters that
are consistent with those used for our numerical simulation: $\log M_1
= 11.24$, $\log L_{0} = 9.95$, $\gamma_1 = 3.18$, $\gamma_2 = 0.245$,
$\sigma_{\rm c} = 0.157$, $\alpha_{\rm s}=-1.18$, $b_0 = -1.17$ , $b_1
= 1.53$, and $b_2 = -0.217$.

We populate the dark mater halos in our simulation with mock galaxies
with luminosities $\log (L/h^{-2}{\rm L}_\odot) \ga 7.0$ using the
following approach. First, each halo is assigned a central galaxy
whose luminosity is drawn from the log-normal distribution of
Eq.\,(\ref{eq:phi_c}). The central galaxy is assumed to be located at
rest at the center of the corresponding halo. Next, we populate the
halo with satellite galaxies via the following steps: (1) obtain the
mean number of satellite galaxies according to the integration of
Eq.(\ref{eq:phi_s}) with luminosities $\log L \ga 7.0$; (2) draw the
actual number of satellite galaxies for the halo in question from a
Poisson distribution with the mean obtained in step (1); (3) assign a
luminosity to each of these satellite galaxy according to
Eq.(\ref{eq:phi_s}). Note that satellite galaxies are allowed to be
brighter than their central galaxy.   Finally the phase-space
  coordinates (positions and velocities) of the satellite galaxies are
  drawn from the randomly selected dark matter particles in the
  halos. As we have tested, populating satellite galaxies in
  phase-space according to an NFW profile yield quite similar
  results.
\begin{figure}
\includegraphics[width=0.5\textwidth]{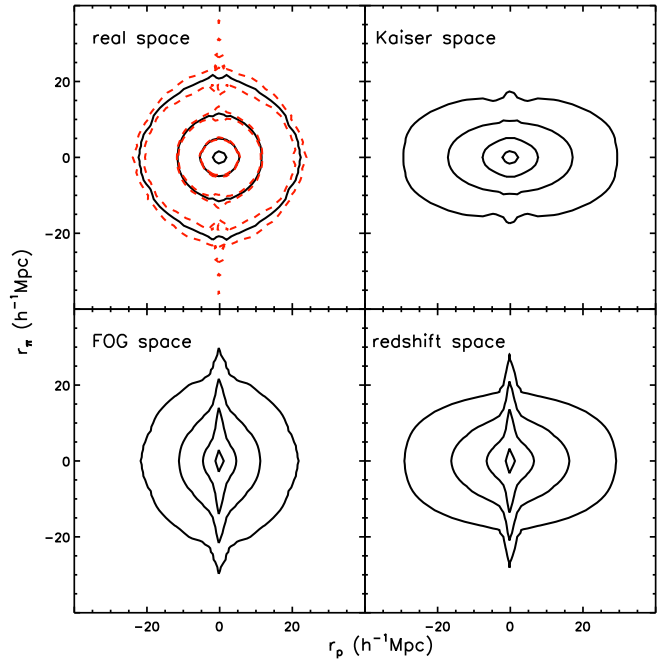}
\caption{The two-dimensional 2PCFs for mock galaxies with absolute
  $r$-band magnitudes in the range $-21\le \rmag \le -20$ for four
  different spaces (see Table~\ref{spaces}): real space (upper left),
  Kaiser space (upper right), FOG space (lower left) and redshift
  space (lower right).  Black contours indicate the average values
  inferred from 10 mock samples. The contour levels correspond to
  $\xi=5,~1,~0.3,~0.1$. The red, dashed contours in the upper
  right-hand panel indicate the $\pm 1\sigma$ cosmic variance.}
\label{fig:2d2pcf}
\end{figure}

\begin{figure}
\center
\includegraphics[width=0.5\textwidth]{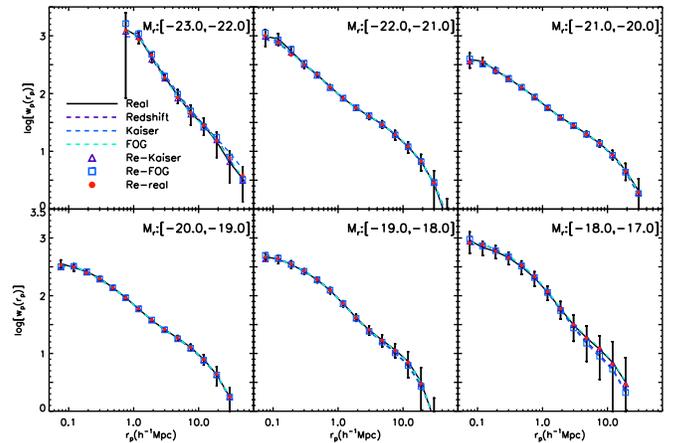}
\caption{Comparison of the projected two-point correlation functions
  in all seven mock spaces. Different panels correspond to different
  bins in absolute $r$-band magnitude, as indicated.  For clarity, the
  error bars, which are obtained from the 10 mock samples, are only
  plotted for the real space results. Note that, as expected, all
  projected correlation functions are virtually
  indistinguishable.}\label{fig:wrp2}
\end{figure}  
\begin{figure}
\includegraphics[width=0.5\textwidth]{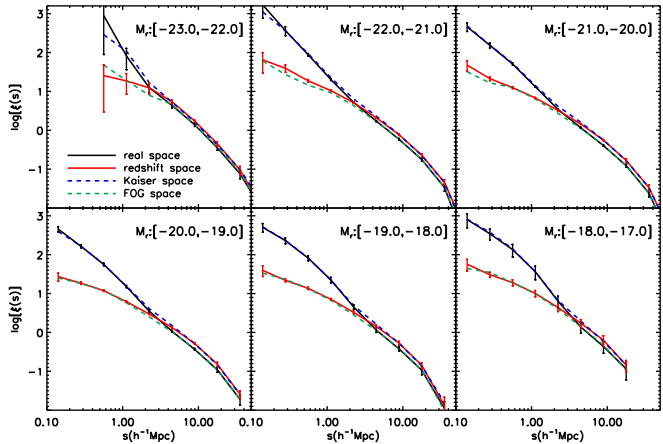}
\caption{The two-point correlation functions of mock galaxies in
  different true spaces. Results are shown for six different
  intervals in absolute $r$-band magnitude, as indicated. For clarity,
  we only plot error bars (expressing the variance among our 10 mock
  samples) for the real and redshift space
  results.}\label{fig:xiS_dif}
\end{figure} 

\begin{figure*}
\center
\includegraphics[width=0.95\textwidth]{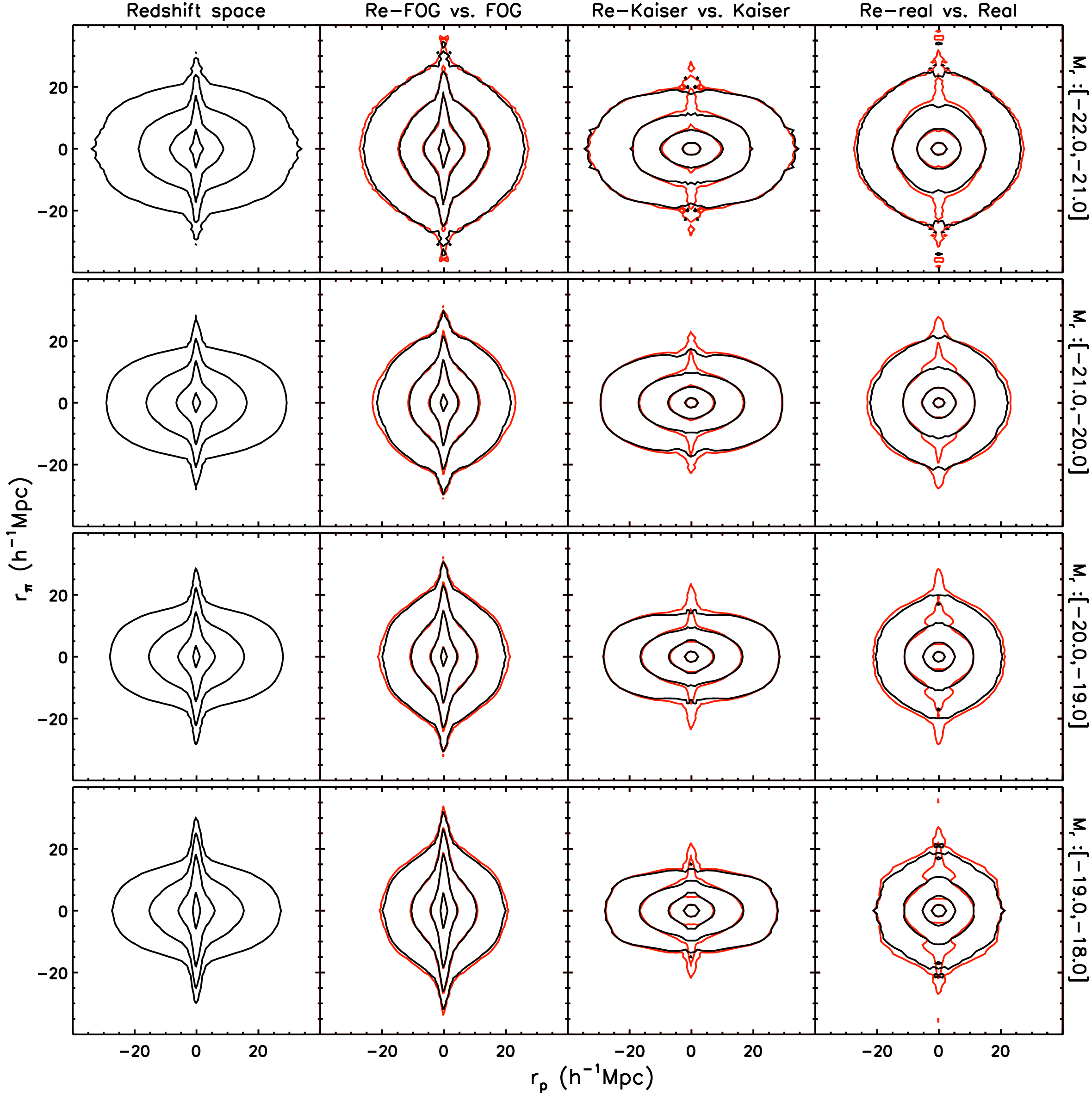}
\caption{Comparison of two-dimensional 2PCFs of mock
  galaxies. Different rows correspond to mock galaxies in different
  absolute $r$-band magnitude bins, as indicated at the right of each
  row. Different columns correspond to different spaces, as indicated
  at the top of each column. Black and red contours correspond to the
  results in the true and reconstructed spaces, respectively, with
  contour levels corresponding to $\xi=5,~1,~0.3,~0.1$. }
\label{fig:2d2pcf2}
\end{figure*}

Next, we proceed to construct mock galaxy samples that have the same
survey selection effects as the SDSS DR7. We stack $3 \times 3 \times
3$ replicas of the populated simulation box and place a virtual
observer at the center of central box. We define a $(\alpha, \delta)$
coordinate system, and remove all mock galaxies that are located
outside of the SDSS DR7 survey region.  We then assign each galaxy the
redshift and $r$-band apparent magnitude according to its distance,
line-of-sight velocity, and luminosity, and select galaxies according
to the position-dependent magnitude limit. Finally, we mimic the
position-dependent completeness by randomly sampling each galaxy using
the completeness masks provided by the SDSS DR7. In order to have an
rough estimation of the cosmic variance, we construct a total of 10
such mock samples by randomly rotating and shifting the boxes in the
stack.  Note that in order to get more accurate estimation of the
  cosmic variance, many more mocks are needed.  From each mock
sample, 6 flux limited (and volume limited) subsamples are constructed
using the redshift and absolute magnitude ranges listed in
Table~\ref{subsamp}.

Finally, in order to disentangle the various redshift distortions, for
each mock galaxy redshift catalog we construct four different versions
that only differ in the redshift $z_{\rm obs}$, assigned to each mock
galaxy: a real-space version in which $z_{\rm obs} = z_{\rm cos}$, a
Kaiser-space version in which $z_{\rm obs} = z_{\rm Kaiser}$
(Eq.~[\ref{zkaiser}]), a FOG-space version in which $z_{\rm obs} =
z_{\rm FOG}$ (Eq.~[\ref{zfog}]), and a redshift-space version in which
$z_{\rm obs}$ is given by Eq.~(\ref{zobs}).
    
\subsection{Results for mock catalogs}

In order to gauge the impact of the various redshift distortions, we
now carry out clustering analyses of the various mock galaxy catalogs
described above.  We start our investigation by computing the
two-dimensional 2PCF, $\xi(r_\rmnp,r_\pi)$. Figure~\ref{fig:2d2pcf}
shows the average results (black solid lines) from the 10 mock samples
for the four true spaces. Here we only show the results for the
$\bm{[}-21.0,-20.0\bm{]}$-subsample, but note that the results for the
other subsamples are qualitatively very similar. The red dashes lines
in the upper-left panel show the $\pm 1\sigma$ cosmic variance as
inferred from our 10 mock samples.  For enhanced clarity, we only show
these in real space. Note that the variance causes small fluctuations
at small transverse separations, $r_\rmnp$, especially at larger
line-of-sight separations, $r_\pi$.

Clearly, the shape of the two-dimensional correlation function is very
different in different spaces: whereas $\xi(r_\rmnp,r_\pi)$ is
isotropic in real space, it is squashed along the line-of-sight on
large scales in Kaiser space, and elongated along the line-of-sight on
small scales in FOG space.  Finally, in redshift space $\xi(r_{\rm
  p},r_\pi)$ reveals the characteristics of both Kaiser and FOG
space. All of this is well known since the seminal work by
\citet{Dav1983}.

Since redshift distortions only displace galaxies along the
line-of-sight, they should not affect the projected correlation
function, $w_\rmnp(r_\rmnp)$, modulo RRSDs that arise from the use
of a finite integration range (see discussion in \S\ref{sec:2pf}). The
lines in Fig.~\ref{fig:wrp2} show the projected 2PCFs in all four true
spaces, and for all six absolute magnitudes bins:
$\bm{[}-23.0,-22.0\bm{]}$, $\bm{[}-22.0,-21.0\bm{]}$,
$\bm{[}-21.0,-20.0\bm{]}$, $\bm{[}-20.0,-19.0\bm{]}$,
$\bm{[}-19.0,-18.0\bm{]}$, $\bm{[}-18.0,-17.0\bm{]}$.  Error bars
reflect the $\pm 1 \sigma$ variance among the 10 mock samples, and,
for clarity, are only plotted for the real space results (they are
very similar in all other spaces).  As expected, the various $w_{\rm
  p}(r_\rmnp)$ are in good agreement with each other, indicating
that the impact of RRSDs is small compared to cosmic variance errors.

Finally, Fig.~\ref{fig:xiS_dif} shows the two-point correlation
function, $\xi(s)$, for the same magnitude bins and the same four
spaces. As before, error bars are obtained from the 10 mock samples,
and only plotted for the real and redshift spaces for clarity.  Unlike
the projected correlation function, $\xi(s)$, clearly reveals the
impact of redshift distortions.  Compared to the real space
correlation function, the $\xi(s)$ in Kaiser space is significantly
boosted at large scales due to the large-scale flows toward
over-dense regions (Kaiser effect). On small scales, however, the
Kaiser space correlation function is virtually indistinguishable from 
the real space correlation function. The $\xi(s)$ in FOG space, on the
other hand, is identical to the real space $\xi(s)$ on large scales,
but dramatically suppressed on small scales. And finally, the $\xi(s)$
in redshift space clearly reveals redshift distortions from both the Kaiser
effect and the FOG effect. 
\begin{figure*}
\center
\includegraphics[width=1.06\textwidth]{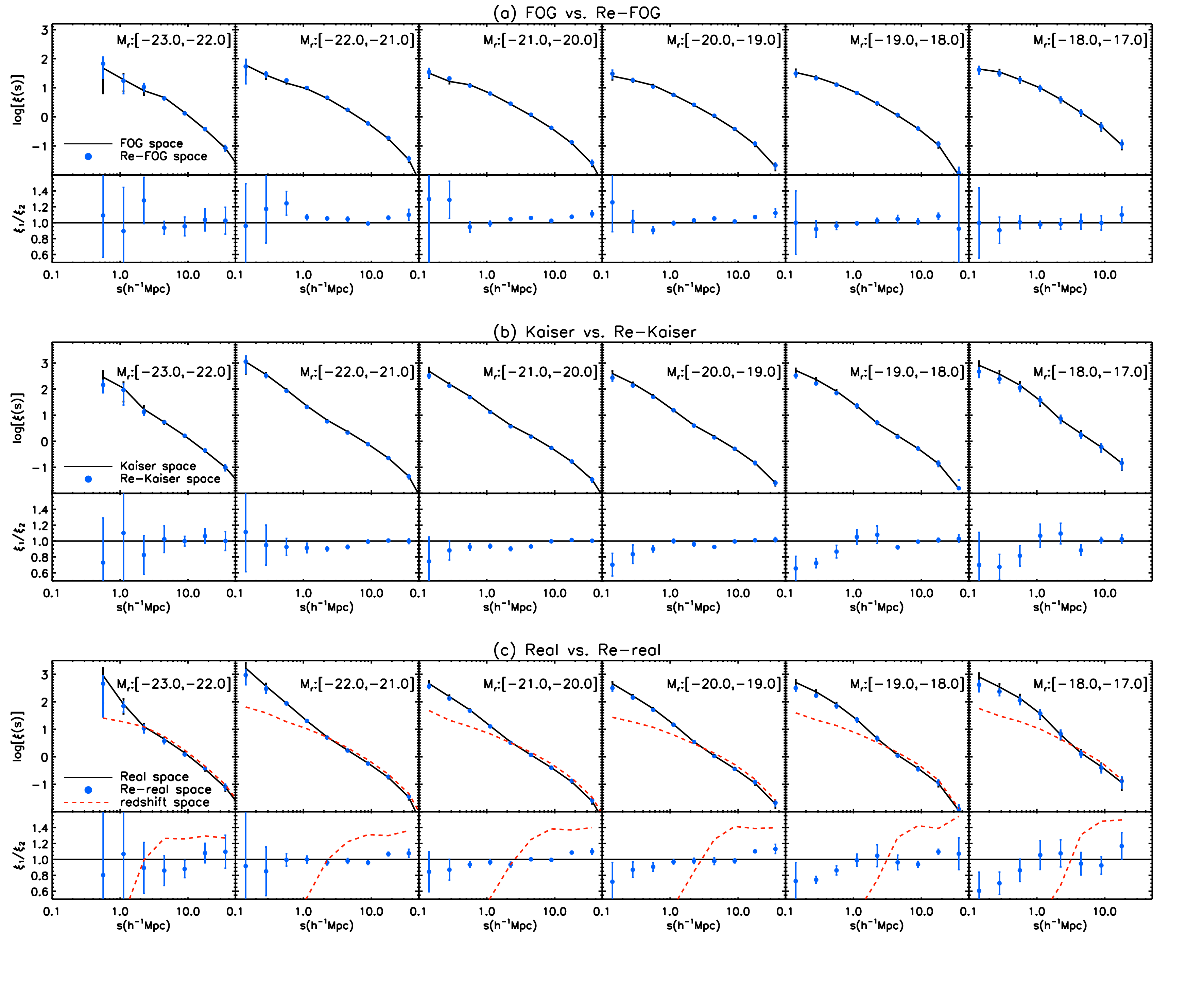}
\caption{2PCFs (upper panels) and 2PCF ratios (lower panels) for mock
  galaxies in (a) FOG vs. re-FOG space, (b) Kaiser vs. re-Kaiser space
  and (c) real vs. re-real space. The solid line in the upper panels
  indicates the 2PCF in the true space, averaged over 10 mock samples,
  while the solid blue circles indicate the corresponding average 2PCF
  in the reconstructed space, with the error bars indicating the $\pm
  1 \sigma$ variance among the 10 mock samples. The lower panels plot
  the average and $\pm 1 \sigma$ variance of the ratio of the 2PCFs in
  the reconstructed space over that in the true space.  For
  comparison, the red dashed lines in the lower panels of (c) indicate
  the ratio of the redshift-space 2PCF to the true real space
  one. Different columns correspond to different $r$-band magnitude
  bins, as indicated.}\label{fig:xiS2}
\end{figure*}

\begin{figure}
\center
\includegraphics[width=0.5\textwidth,height=0.25\textwidth]{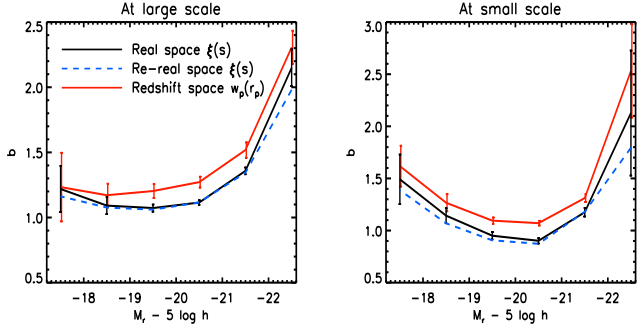}
\caption{The bias factor for mock galaxies as a function of galaxy
  absolute magnitude in real space (black solid lines), re-real space
  (blue dashed lines), and redshift space (red solid lines). The bias
  factors for real and re-real space are defined as the ratios of the
  measured $\xi(s)$ to that of the dark matter over the ranges of
  $4\mpch < s < 20\mpch$ (left panel) and $0.5\mpch < s < 2\mpch$
  (right panel). The bias factor for redshift space is defined as the
  ratios of $w_p(r_p)$ between galaxies and dark matter over the range
  $4\mpch < r_p < 20 \mpch$ (left panel) and $0.5\mpch <r_p < 2 \mpch$
  (right panel). Here the integration limit, $r_{max}$, in computing
  $w_p(r_p)$ from $\xi(r_{\rm p}, r_{\pi})$, is set to be
  $60\mpch$. The error bars, shown only for real space and redshift
  space, correspond to $1\sigma$ variance among 10 mock samples.
}\label{fig:bias}
\end{figure}       

\begin{figure}
\center
\includegraphics[width=0.5\textwidth]{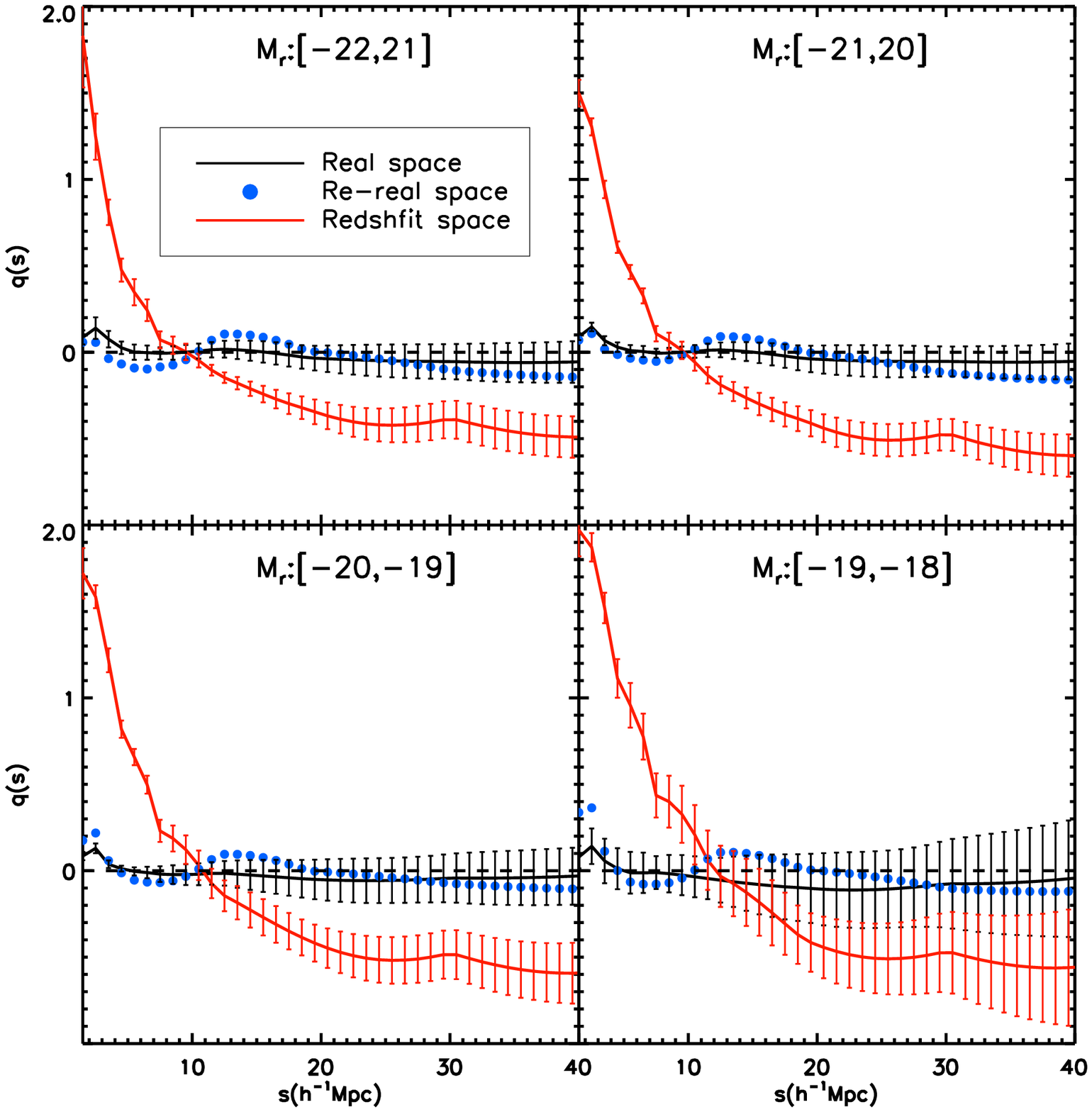}
\caption{The quadrupole-to-monopole ratio $q(s)$ for mock galaxies in
  real space (black solid lines), in reconstructed real space (blue
  solid circles) and in redshift space (red solid lines). Results
  indicate the averages obtained from 10 mocks, with error bars (for
  clarity, not plotted for re-real space) indicating the corresponding
  $\pm 1\sigma$ variance. Different panels correspond to different
  bins in absolute $r$-band magnitude, as indicated.}\label{fig:qs}
\end{figure}       

\subsection{Results for reconstructed catalogs}
\label{sec:recres}

Thus far we have constructed mock SDSS DR7 galaxy catalogs in four
true spaces that allow us to disentangle the impact of the FOG effect
on small scales from the Kaiser effect on large scales. We have shown
that the results from statistical analyses of galaxy clustering in
these different spaces agree with expectations. We now proceed with
using these mock catalogs to test the reliability and accuracy of the
reconstruction method described in \S\ref{sec:correction}.  We start
by running the halo-based group finder of \citet{Yang2005, Yang2007}
over each of the separate true space mock galaxy catalogs.  This
yields corresponding mock group catalogs, in which each group is
assigned a halo mass based on its characteristic luminosity, as
described in Y07. Similar to the SDSS group catalog, the mock group
catalogs are also complete to $z \sim 0.12$ for groups with an
assigned halo mass $M_\rmh \geq 10^{12.5} \msunh$.  We thus
adopt a threshold mass of $M_{\rm th} = 10^{12.5} \msunh$ and restrict
our reconstruction to the volume covering the redshift range $0.01
\leq z \leq 0.12$.

Next we use the redshift distortion correction method described in
\S\ref{sec:correction} to obtain mock galaxy catalogs in re-FOG,
re-Kaiser and re-real space.  In this subsection we focus on comparing
the clustering of galaxies in the reconstructed spaces with that in
the corresponding true spaces.  The goal is to investigate the
accuracy with which the reconstruction method can recover the
distribution of galaxies in real space. Throughout we characterize the
clustering using the various two-point correlation functions
introduced above and we use the 10 independent mock samples to gauge
the impact of sample variance.

\subsubsection{The two-dimensional correlation function $\xi(r_\rmnp,r_\pi)$}
\label{sec:2DCFtest}

We start with a qualitative, visual comparison based on the
two-dimensional 2PCF $\xi(r_\rmnp,r_\pi)$. Different rows in
Fig.~\ref{fig:2d2pcf2} correspond to different magnitude bins, as
indicated at the right-hand side of each row. From left to right, the
different columns show the results in redshift space, a comparison of
FOG vs. re-FOG, a comparison of Kaiser vs. re-Kaiser, and a comparison
of real vs. re-real. In each case black and red contours correspond to
the true and reconstructed spaces, respectively.

The $\xi(r_\rmnp,r_\pi)$ in redshift space is clearly anisotropic,
revealing fingers-of-God on small scales and the impact of the Kaiser
effect on large scales.  After correcting for the Kaiser effect, the
resulting $\xi(r_\rmnp,r_\pi)$ in re-FOG space is clearly more
isotropic on large scales. As expected, it still reveals the impact of
the FOG effect, which distort the contours from being perfectly round.
A comparison with the $\xi(r_\rmnp,r_\pi)$ in FOG space shows that
the correction for the Kaiser effect is overall very successful,
except for small differences in the outer contour (corresponding to
$\xi = 0.1$).

Comparing the $\xi(r_\rmnp,r_\pi)$ in re-Kaiser space (third column
from the left) with that in redshift space (left-hand column) shows
that our method of FOG compression is fairly accurate. However, a
comparison with the true Kaiser space results (black contours in third
column) reveals that the method is not perfect.  On small scales, the
$\xi(r_\rmnp,r_\pi)$ in re-Kaiser space shows very nice agreement with
that in the real space.  On large scales, however, the correlation
function in re-Kaiser space reveals residual FOG effects.  These
shortcomings of the FOG compression may arise from problems with the
group finder, including errors in group membership determination
(`fracturing' and `fusing' of groups), errors in the designation of
centrals and satellites, and errors in the halo mass assignment. These
errors are characteristic of all group finders, and are virtually
impossible to avoid \citep[see][for details]{Cam2015}.

Finally, the results in the rightmost column show that the
reconstruction of $\xi(r_\rmnp,r_\pi)$ in real space manifests both
the problems with the Kaiser correction and the FOG compression.
Overall, though, comparing the correlation function in re-real space
with that in redshift space, it is clear that the reconstruction
method has successfully corrected for the majority of redshift space
distortions.  In order to make this more quantitative, we now focus on
$\xi(s)$.
\begin{figure*}
\center
\includegraphics[width=0.9\textwidth]{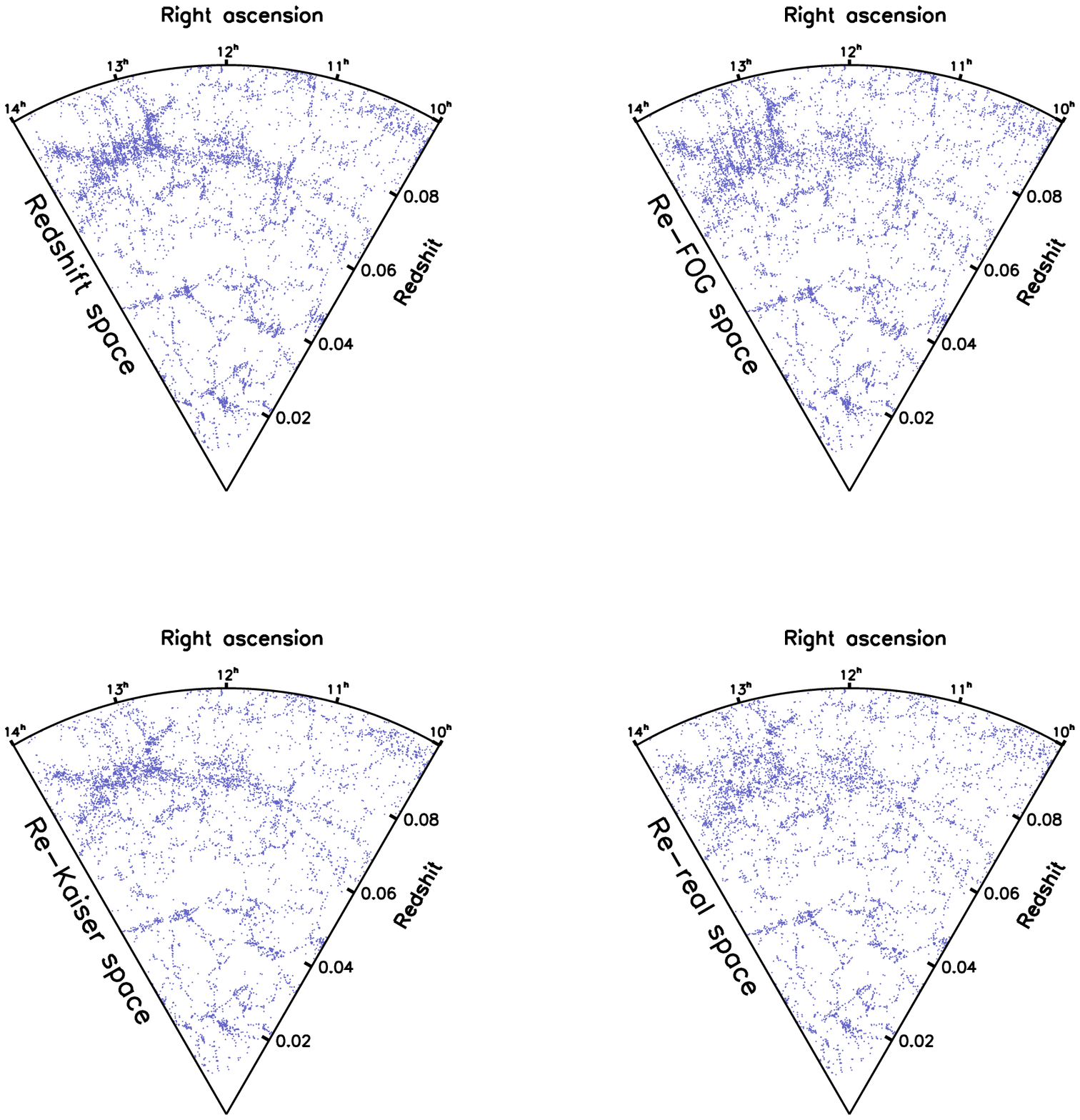}
\caption{Distribution of a subset of SDSS DR7 galaxies in a
  $4^{\circ}$ slice with right ascensions in the range from
  $14^{\rmh}$ to $10^{\rmh}$ and with redshifts $0.01 \leq z
  \leq0.1$. Clockwise from the top-left, the panels show results in
  redshift space, in reconstructed FOG space, in reconstructed real
  space, and in reconstructed Kaiser space. Note how the Sloan Great
  Wall, evident in the upper left corner of each panel, is far less
  pronounced in real space than in redshift space.}\label{fig:xy_dr7}
\end{figure*}

\subsubsection{The one-dimensional correlation function $\xi(s)$}
\label{sec:1DCFtest}

Figure~\ref{fig:xiS2} compares the 2PCF, obtained by averaging results
from all 10 mocks, in a true space ($\xi_{\rm true}$, solid lines) to
that in the corresponding reconstructed space ($\xi_{\rm recon}$, blue
filled circles).  From top to bottom, the three parts of this figure
show a comparison of (a) FOG space versus re-FOG space, (b) Kaiser
space versus re-Kaiser space, and (c) real space versus re-real space.
Different columns correspond to different magnitude bins, as
indicated, and error bars indicate the variance among the 10 mock
samples. In each part, the upper panels show the actual 2PCFs, while
the lower panels plot $\xi_{\rm recon}/\xi_{\rm true}$\footnote{Note
  that we plot the average of the ratios, rather than the ratio of the
  averages}. Overall, the correlation functions in the reconstructed
spaces are in excellent agreement with those in their corresponding
true spaces, with the vast majority of data points being consistent
with $\xi_{\rm recon}/\xi_{\rm true}=1$ within $1 \sigma$. Recall that
$\sigma$ reflects the measurement error due to cosmic variance in a
SDSS-like survey.

As is evident from the middle part (b), the FOG compression seems to
systematically under predict the Kaiser-space 2PCF for faint galaxies.
 The effect, which results from inaccuracies in the group finder,
  is somewhat significant in the two low mass bins. Thus in an
  accurate modeling for the halo occupation distribution of galaxies
  for these faint galaxies, one needs to taken this effect into
  account.  For brighter galaxies, over the range of scales $0.2 \leq
  (s/\mpch) \leq 20$, the average values of $\xi_{\rm
    re-real}/\xi_{\rm real}$ is $1.00 \pm 0.050$.  Hence, we conclude
  that over those scales the reconstruction of the real space
  correlation function is accurate at five percent level. For
comparison, the dashed lines in the bottom part (c) of
Fig.~\ref{fig:xiS2} correspond to the 2PCF in redshift space.  On
small scales ($r < 1 \mpch$), the clustering strength in redshift
space is suppressed by $ \sim70\%$ on average, compared to that in real
space.  On large scales, ($r > 2 \mpch$) it is boosted by $\sim30\%$ on
average.

\subsubsection{The projected correlation function $w_\rmnp(r_\rmnp)$}
\label{sec:wprptest}

Moreover, since our reconstruction only `displaces' galaxies along the
line-of-sight, the reconstruction method has no impact on the
projected correlation function, $w_\rmnp(r_\rmnp)$, other than
scattering a few galaxy pairs in and out of the sample due to the
finite integration range used ($\vert r_\pi \vert \leq 100
\mpch$). This effect is entirely negligible, though, as is evident
from Fig.~\ref{fig:wrp2}, which shows the results for all of our seven
spaces (four true space and three reconstructed spaces). There are no
significant differences among these different projected correlation
functions.
\begin{figure*}
\center
\includegraphics[width=0.9\textwidth]{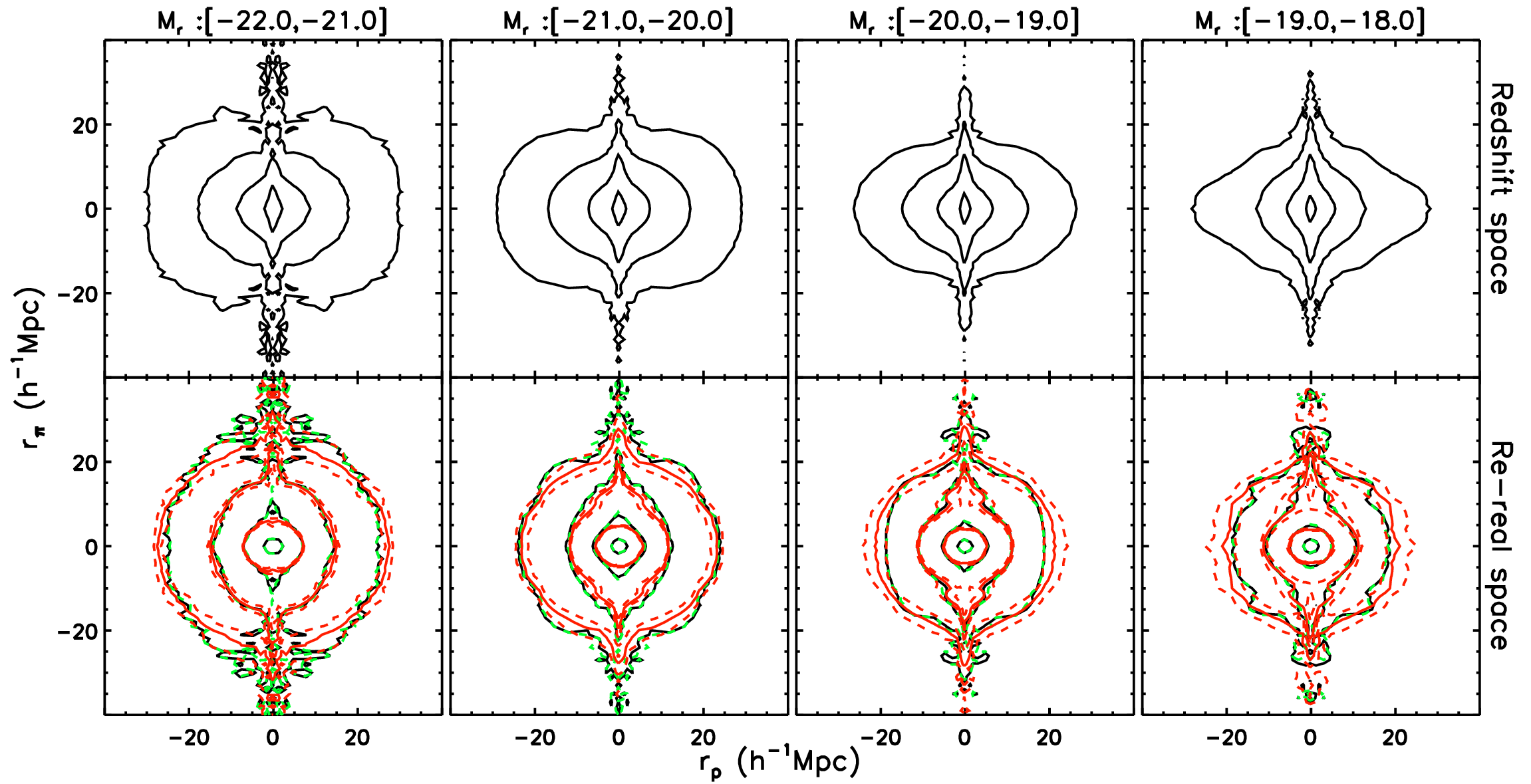}
\caption{The two-dimensional 2PCFs for SDSS DR7 data in redshift space
  (upper panels) and re-real space (lower panels). Different columns
  correspond to different magnitude bins as indicated.  The black
    solid lines in the lower panels are for WMAP9 cosmology, and the
    green dashed lines are for Planck cosmology. The red, solid and
  dashed contours indicate the average and $\pm 1 \sigma$ variance,
  respectively, obtained from the 10 mock galaxy samples in re-real
  space. Contour levels correspond to $\xi = 5,~1,
  ~0.3,~0.1$.}\label{fig:2d2pcf_dr7}
\end{figure*}       

\subsubsection{The bias factor}
\label{sec:btest}

The correlation function of galaxies relative to that of dark matter 
is usually described by a bias factor, which is defined 
\begin{equation}
\xi_{gg}(s)=b^2\xi_{mm}(s)\,,
\end{equation}
where $\xi_{gg}$ and $\xi_{mm}$ are the correlation functions of
galaxies and mass, respectively. In general, the bias factor $b$ may
depend on $s$.  Figure \ref{fig:bias} shows the best-fitting bias
factor, as a function of galaxy luminosity, obtained from the measured
$\xi(s)$ for mock galaxies relative to the correlation function of
dark matter at $z=0.1$. The real-space and reconstructed real-space
$b$ shown in the left panel are obtained from using the values of
$\xi(s)$ at large scales, $4\mpch < s < 20\mpch$, while in the right
panels they are obtained using the correlation functions on small
scales, $0.5\mpch < s < 2\mpch$.  For comparison, we also show in
Figure \ref{fig:bias} the bias factor based on the projected 2PCFs
(red lines), defined as the ratios of $w_p(r_p)$ between galaxies and
dark matter over the range $4\mpch < r_p < 20 \mpch$ (left panel) and
$0.5\mpch <r_p < 2 \mpch$ (right panel). As one can see, the
reconstructed real-space $b$ closely matches that in the real-space,
while the traditional method based on $w(r_p)$ leads to larger errors
and biased results relative to the true real-space values.

\subsubsection{The quadrupole-to-monopole ratio $q(s)$}
\label{sec:qstest}

As a final diagnostic of our reconstruction performance, we consider
the quadrupole-to-monopole ratio, which is defined as
\begin{equation}
q(s) \equiv \frac{\xi_2(s)}
    {\frac{3}{s^3}\int^s_0{\xi_0(s') \, s'^2 \, {\rm d}s'} - \xi_0(s)} 
\end{equation}
with $\xi_l(s)$ given by
\begin{equation}
\xi_l(s)=\frac{2l+1}{2}\int^1_{-1}{\xi(r_\rmnp,r_\pi) \, \mathscr{P}_l(\mu) \, {\rm d}\mu}
\end{equation}
  where $\mathscr{P}_l(\mu)$ is the $l$th Legendre polynomial.
In redshift space, the Kaiser effect causes the quadrupole-to-monopole
ratio to become negative on large scales, asymptoting towards
\begin{equation}
  q(s) = {-{4 \over 3}\beta - {4 \over 7}\beta^2 \over
    1+{2\over 3}\beta + {1 \over 5}\beta^2}\,,
\end{equation}
where $\beta = f(\Omega)/b$ with $b$ the bias parameter of the galaxy
population under consideration \citep[e.g.,][]{Ham1992, Col1994}.  On
small scales the FOG effect causes $q(s)$ to become positive. In real
space, however, we expect isotropy to results in a quadrupole
$\xi_2(s) = 0$.  Hence, if the correction for redshift distortions is
successful, the resulting clustering should have a vanishing
quadrupole, and thus $q(s) = 0$.

Figure~\ref{fig:qs} shows the quadrupole-to-monopole ratio in our real
space, re-real space and redshift space mocks. Different panels
correspond to different magnitude bins, as indicated. As expected, in
redshift space $q(s)$ has large deviations from zero on both small and
large scales, while in real space $q(s)$ is close to zero (except for a small
positive signal for $r \lta 3 \mpch$, which is due to noise). In
re-real space, the quadrupole-to-monopole ratio in the re-real space
is consistent with zero within the error bars on large scales ($\gta
12 \mpch$). On smaller scales, all magnitude bins reveal a slightly
negative $q(s)$.  This is a consequence of the over-correction for the
FOG effect on small scales discussed in \S\ref{sec:2DCFtest}
(cf. Fig.~\ref{fig:2d2pcf2}), which has its origin in inaccuracies
associated with the galaxy group finder.


\section{APPLICATION TO THE SLOAN DIGITAL SKY SURVEY}
\label{sec:apply}
 
Based on the analyses of the mock galaxy samples discussed in
\S\ref{sec:mock}, we conclude that our reconstruction method can
accurately correct for redshift space distortions in a statistical
sense. In this section we apply exactly the same method to the SDSS
DR7.  As described in \S\ref{sec:data} we follow W12 and reconstruct
the velocity field on quasi-linear scales using the mass distribution
reconstructed from galaxy groups of Y07 in the redshift range $0.01
\leq z \leq 0.12$ and with assigned halo masses $\log(M_\rmh/\msunh)
\geq 12.5$. We use the velocities derived to correct for the Kaiser
effect using the method described in \S\ref{sec:KaiserCORR}.  Finally,
we correct for the FOG effect by assigning all galaxies new positions
within their groups based on the method described in
\S\ref{sec:FOGCORR}. We apply this method to all the 396,068 galaxies
in the NGC region. The reconstructed real space galaxy catalog is
publicly available through
\href{url}{http://gax.shao.ac.cn/data/data1/SDSS7\_REAL.tar}.

\subsection{The galaxy distribution}
\label{sec:distribution}

To visualize the effects of our reconstruction method on galaxy
distribution, we shown in Fig.~\ref{fig:xy_dr7} the distributions of
galaxies with declination $|\delta| < 4^{\circ}$, right ascensions
$10^\rmh \leq \alpha \leq 14^\rmh$, and redshifts $0.01 \leq z \leq
0.1$. The four different panels show the galaxy distributions in
redshift space (upper-left panel), re-FOG space (upper-right panel),
re-Kaiser space (lower-left panel), and re-real space (lower-right
panel), respectively.  Note that the volume chosen includes the Sloan
Great Wall, which is readily visible in the upper left corner ($z \sim
0.085$ and $12^\rmh \leq \alpha \leq 14^\rmh$).

There are a few noteworthy trends. First of all, the prominent
`finger' structures clearly visible in redshift space are no longer
visible in the re-Kaiser space, indicating that our FOG compression is
successful.  Comparing the redshift space distribution with that in
re-real space, one sees that the distribution in re-real space appears
more diffused on large scales, more compressed on small scales.  In
particular, the Sloan Great Wall is clearly much broader, and thus
less pronounced, in the re-FOG and re-real spaces. This suggests that
the Sloan Great Wall is not as dominant an over-dense structure as it
appears to be in redshift space, but that its apparent over-density is
strongly enhanced by the Kaiser effect.

It is also clear from Fig.~\ref{fig:xy_dr7} that some geometrical
properties of the large scale structure may also be affected as one
goes from real space to redshift space distortion. For example, the
voids appear to be smaller and the filamentary structures less
prominent in real space.  Clearly, detailed analyses are needed in
order to quantify the effects, and our reconstructed real-space
catalog of SDSS DR7 provides a unique resource for such studies.
\begin{figure*}
\center
\includegraphics[width=0.75\textwidth,height=0.9\textwidth]{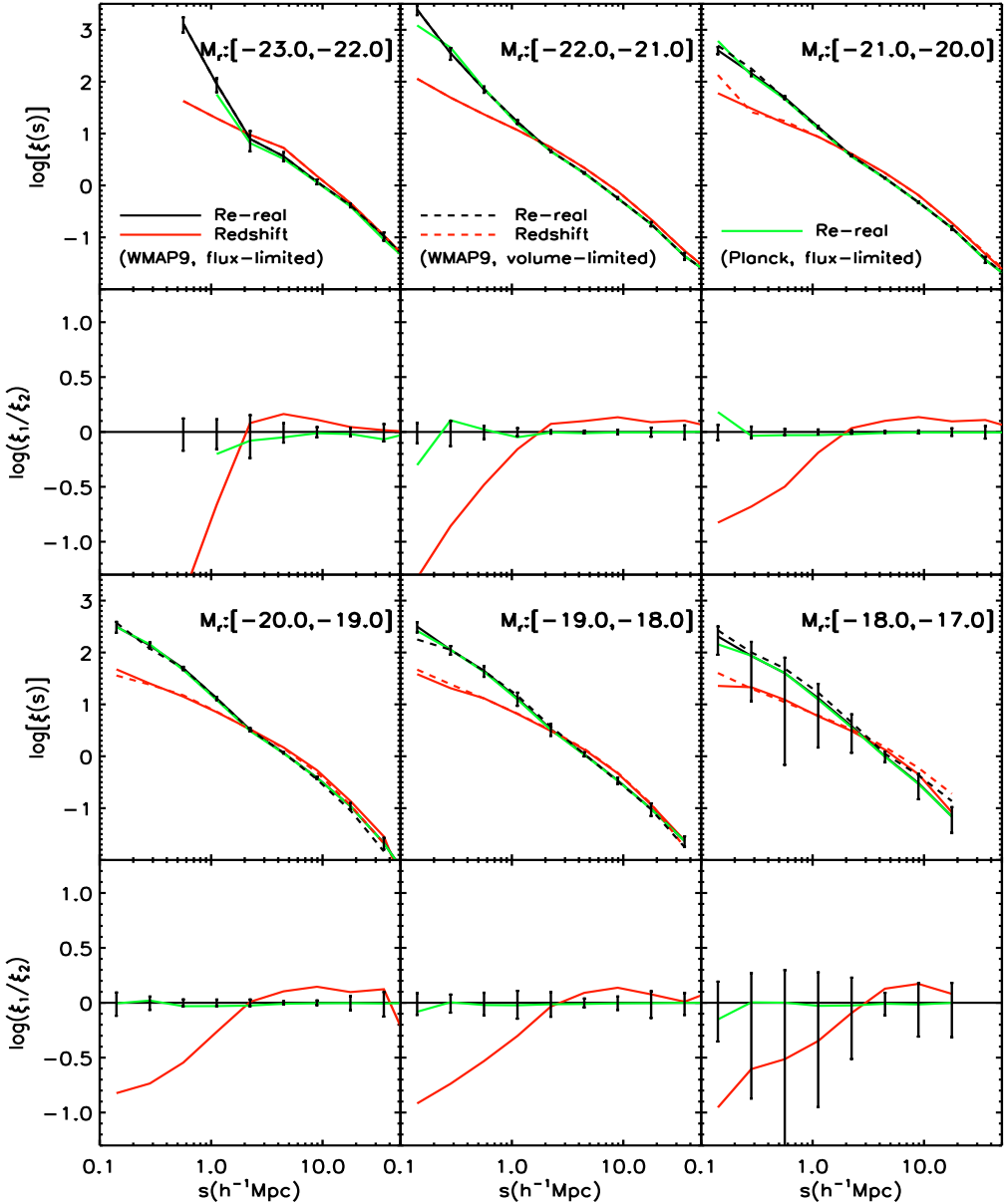}
\caption{The 2PCFs and 2PCF ratios for SDSS galaxies in redshift space
  (red lines) and the reconstructed real space ( WMAP9 with black
    lines, Planck with green lines). For comparison, the 2PCFs of the
  flux-limited (solid lines) and volume-limited samples (dashed lines)
  are both shown.  Error bars, only shown for the re-real space
  results, are the $\pm 1\sigma$ variance among the 10 mock samples
  discussed in \S\ref{sec:mockcat}. The red curves in the lower panels
  are the ratios of the redshift space to re-real space
  2PCFs. Different columns correspond to different bins in absolute
  $r$-band magnitude, as indicated.}
  \label{fig:xiS_dr7}
\end{figure*}  

\begin{figure*}
\center
\includegraphics[width=0.78\textwidth]{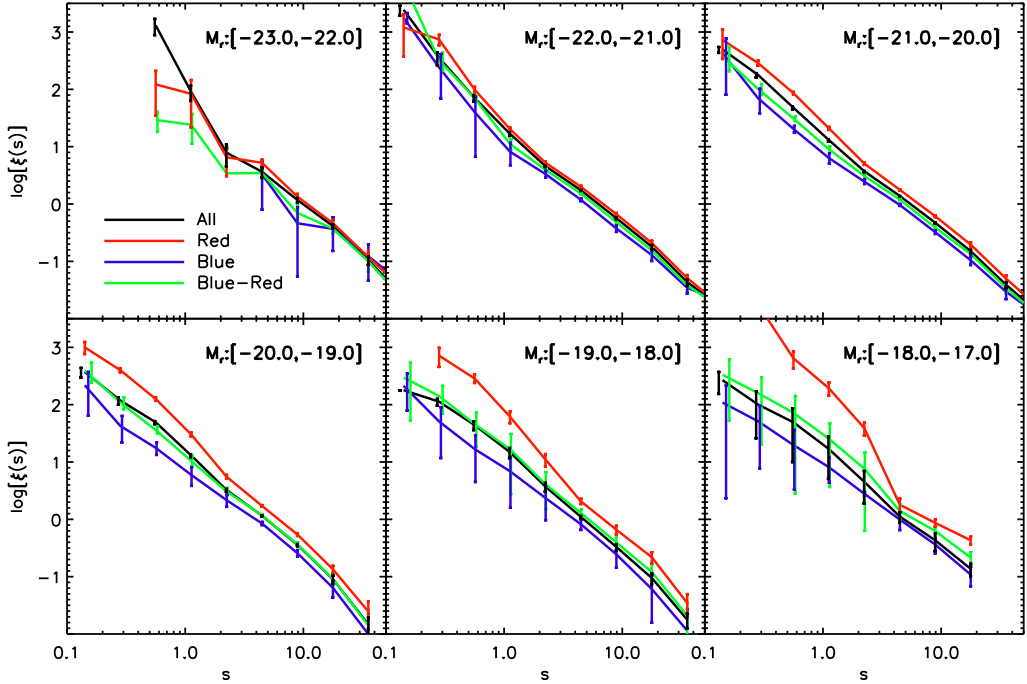}
\caption{ The 2PCFs of red and blue galaxies for SDSS volume-limited
  samples in re-real space. Different columns correspond to different
  bins in the absolute $r$-band magnitude, as indicated. Red and blue
  lines show the autocorrelation functions of the red and blue
  subsamples, respectively, while black lines show the autocorrelation
  of the full sample in each $M_r$ bin. Green lines show the
  cross-correlation function between red and blue
  galaxies.}\label{fig:xiS_color}
\end{figure*}  
\begin{table*}
\center
\scalebox{0.8}{
\begin{threeparttable}[c]
  \caption{The 2PCFs obtained from SDSS DR7 in redshift space and reconstructed real space. }\label{2PCF_dr7}
\setlength{\tabcolsep}{4.0pt}
\begin{tabular}{rrrrrrrrrrrrrrrrrr}
\toprule
\multicolumn{2}{c}{~}  &
\multicolumn{2}{c}{\multirow{2}{*}{$~~~~~\bm{[}-23,~-22\bm{]}$}} & \multicolumn{2}{c}{\multirow{2}{*}{$~~\bm{[}-22,~-21\bm{]}$}} &
\multicolumn{3}{c}{\multirow{2}{*}{$~~~~~\bm{[}-21,~-20\bm{]}$}} & \multicolumn{3}{c}{\multirow{2}{*}{$~~\bm{[}-20,~-19\bm{]}$}} &
\multicolumn{3}{c}{\multirow{2}{*}{$\bm{[}-19,~-18\bm{]}$}} & \multicolumn{3}{c}{\multirow{2}{*}{$\bm{[}-18,~-17]\bm{}$}} \\\\

 & \multicolumn{1}{c}{~~$r$}   &
   \multicolumn{1}{c}{~~~$\xi(\xi')$}&\multicolumn{1}{c}{~~~$\mathop{\Delta\xi}$} & 
   \multicolumn{1}{c}{~~~$\xi(\xi')$}&\multicolumn{1}{c}{~~~$\mathop{\Delta\xi}$} & 
   \multicolumn{1}{c}{~~$\xi$} &\multicolumn{1}{c}{~~$\xi'$} &\multicolumn{1}{c}{~~~$\mathop{\Delta\xi}$} &
   \multicolumn{1}{c}{~~$\xi$} &\multicolumn{1}{c}{~~$\xi'$} &\multicolumn{1}{c}{~~~$\mathop{\Delta\xi}$} &
   \multicolumn{1}{c}{~~$\xi$} &\multicolumn{1}{c}{~~$\xi'$} &\multicolumn{1}{c}{~~~$\mathop{\Delta\xi}$} &
   \multicolumn{1}{c}{~~$\xi$} &\multicolumn{1}{c}{~~$\xi'$} &\multicolumn{1}{c}{~~~$\mathop{\Delta\xi}$}   \\
~& 0.14& ~& ~&113.849& 34.734& 59.898&134.516& 14.312& 47.269& 36.029&  6.318& 38.225& 46.719& 12.109& 22.684& 40.278& 19.146\\\multicolumn{1}{c}{\multirow{1}{*}{\rotatebox{90}{Redshift space}}}
~& 0.28& ~& ~& 49.288&  8.310& 30.435& 25.849&  2.486& 25.203& 24.042&  1.322& 20.512& 24.432&  1.770& 21.377& 20.144&  4.651\\
~& 0.56& 42.401& 22.572& 23.400&  1.410& 15.645& 17.469&  0.730& 14.101& 14.980&  0.519& 13.083& 12.962&  0.954& 12.261& 11.008&  3.446\\
~& 1.12& 19.626& 10.093& 11.659&  0.382&  8.686&  8.670&  0.231&  7.078&  7.234&  0.226&  6.540&  6.643&  0.356&  5.915&  6.134&  1.894\\
~& 2.24&  9.474&  4.009&  5.424&  0.136&  4.070&  4.044&  0.086&  3.313&  3.275&  0.097&  3.070&  3.174&  0.225&  3.067&  3.298&  0.918\\
~& 4.47&  5.308&  0.645&  2.191&  0.055&  1.750&  1.730&  0.034&  1.479&  1.423&  0.047&  1.346&  1.392&  0.117&  1.339&  1.504&  0.406\\
~& 8.91&  1.516&  0.090&  0.773&  0.032&  0.652&  0.646&  0.022&  0.539&  0.488&  0.030&  0.470&  0.483&  0.062&  0.450&  0.598&  0.195\\
~&17.78&  0.458&  0.058&  0.221&  0.022&  0.190&  0.187&  0.014&  0.137&  0.114&  0.020&  0.116&  0.122&  0.024&  0.083&  0.190&  0.046\\
~&35.48&  0.109&  0.023&  0.055&  0.007&  0.048&  0.051&  0.006&  0.028&  0.021&  0.006&  0.024&  0.018&  0.008&  ~& ~& ~\\
\hline
~& 0.14&  ~&  ~&1386.112& 517.165& 401.879& 496.315&  63.661& 314.918& 372.194&  75.121& 315.778& 178.442&  52.512& 203.865& 267.397&113.507\\\multicolumn{1}{c}{\multirow{1}{*}{\rotatebox{90}{Re-real space}}}
~& 0.28&  ~&  ~& 403.161&  91.274& 145.658& 178.764&  17.368& 137.039& 118.726&  18.983& 112.085& 112.744&  16.460&  85.958& 100.588& 74.667\\
~& 0.56&1308.251&  425.252&  82.973&   9.812&  49.341&  47.769&   3.154&  49.473&  49.278&   3.563&  44.412&  44.125&   8.146&  40.043&  49.205& 39.238\\
~& 1.12&  90.370&  27.875&  16.178&   1.398&  13.415&  13.065&   0.705&  13.028&  13.050&   0.905&  13.116&  14.809&   3.118&  13.187&  16.850& 11.799\\
~& 2.24&   7.865&   3.316&   4.576&   0.161&   3.757&   3.721&   0.132&   3.248&   3.237&   0.229&   3.321&   3.732&   0.687&   3.799&   4.501&  2.614\\
~& 4.47&   3.645&   0.730&   1.737&   0.051&   1.386&   1.384&   0.028&   1.164&   1.162&   0.037&   1.092&   1.103&   0.071&   0.996&   1.116&  0.232\\
~& 8.91&   1.175&   0.125&   0.567&   0.027&   0.478&   0.471&   0.015&   0.385&   0.359&   0.019&   0.342&   0.330&   0.048&   0.303&   0.431&  0.153\\
~&17.78&   0.413&   0.034&   0.180&   0.017&   0.152&   0.150&   0.012&   0.110&   0.090&   0.016&   0.097&   0.095&   0.021&   0.069&   0.137&  0.035\\
~&35.48&   0.105&   0.019&   0.044&   0.006&   0.037&   0.039&   0.005&   0.021&   0.015&   0.005&   0.023&   0.018&   0.005&  ~&  ~& ~\\
\bottomrule
\end{tabular}
\textbf{Notes.} $r$: the comoving distances in units of
$\mpch$.~~$\xi$: the two-point correlation function for flux 
limited samples.~~$\xi'$: the two-point correlation function for 
volume-limited samples (the flux- and volume- limited samples are the 
same for the first two samples).
~~$\mathop{\Delta\xi}$: the $1\sigma$ error of $\xi(s)$
estimated using 10 mock samples.
\end{threeparttable}}
\end{table*}

\subsection{The clustering of galaxies}
\label{sec:clusan}

Next we investigate the galaxy clustering properties.   It is
  important to note that the reconstruction to obtain the re-real
  space is cosmology dependent. The bias parameter $b$, the halo mass
  assignments to galaxy groups, and the distance-redshift relation are
  all cosmology dependent. In the reconstruction of the SDSS-DR7, we
  have adopted the cosmological parameters as inferred from the WMAP9.
  To check the impact of cosmology on our results, we also adopt a
  Planck cosmology ($\Omega_\rmm = 0.308$, $\Omega_{\Lambda} = 0.692$,
  $n_{\rm s}=0.968$, $h=H_0/(100 \kmsmpc) = 0.678$ and $\sigma_8 =
  0.815$) \citep{Planck2015} in our reconstruction. In general, there
  is no large distinction between the results for the two cosmologies.
  In what follows, we mainly focus on the results for the WMAP9
  cosmology, results for Planck cosmology are also presented where
  necessary.

The black contours in Fig.~\ref{fig:2d2pcf_dr7} show the two
dimensional 2PCFs $\xi(r_\rmnp,r_{\pi})$, for galaxies in four
luminosity bins, in redshift space (upper panels) and re-real space
(lower panels) for WMAP9 cosmology.  While the green contours are
  results for Planck cosmology, which show quite good agreement with
  those for WMAP9 cosmology. After the correction of the redshift
distortion, the $\xi(r_{\rm p},r_{\pi})$ is clearly much more
isotropic than in redshift space.  However, it is also clear that the
correction is not perfect, especially on small transverse scales where
residual deviations from isotropy are apparent.  To assess the
significance of these deviations, we use the 10 mock re-real space
samples of \S\ref{sec:recres} to estimate the significance of the
cosmic variance.  The solid and dashed red contours in the lower
panels of Fig.~\ref{fig:2d2pcf_dr7} show the average and $\pm 1\sigma$
variance among these 10 mock samples. Clearly, the variance is large,
and most of the black contours fall within these $1\sigma$ error
ranges, suggesting that the remaining deviations from isotropy are
mainly a manifestation of sampling variance, rather than a systematic
error in the reconstruction method.

\begin{table*}
\center
\scalebox{0.78}{
\begin{threeparttable}[c]
  \caption{The color-dependence of the 2PCF measured from SDSS DR7 in 
  the reconstructed real space. }\label{2PCF_color}
\setlength{\tabcolsep}{6.0pt}
\begin{tabular}{rrrrrrrrrrrrrr}
\toprule
\multicolumn{2}{c}{~}  &
\multicolumn{2}{c}{\multirow{2}{*}{$~~~~~\bm{[}-23,~-22\bm{]}$}} & \multicolumn{2}{c}{\multirow{2}{*}{$~~\bm{[}-22,~-21\bm{]}$}} &
\multicolumn{2}{c}{\multirow{2}{*}{$~~~~~\bm{[}-21,~-20\bm{]}$}} & \multicolumn{2}{c}{\multirow{2}{*}{$~~\bm{[}-20,~-19\bm{]}$}} &
\multicolumn{2}{c}{\multirow{2}{*}{$\bm{[}-19,~-18\bm{]}$}} & \multicolumn{2}{c}{\multirow{2}{*}{$\bm{[}-18,~-17]\bm{}$}} \\\\

 & \multicolumn{1}{c}{~~$r$}   &
   \multicolumn{1}{c}{~~~$\xi$}&\multicolumn{1}{c}{~~~$\mathop{\Delta\xi}$} & 
   \multicolumn{1}{c}{~~~$\xi$}&\multicolumn{1}{c}{~~~$\mathop{\Delta\xi}$} & 
   \multicolumn{1}{c}{~~$\xi$} &\multicolumn{1}{c}{~~~$\mathop{\Delta\xi}$} &
   \multicolumn{1}{c}{~~$\xi$} &\multicolumn{1}{c}{~~~$\mathop{\Delta\xi}$} &
   \multicolumn{1}{c}{~~$\xi$} &\multicolumn{1}{c}{~~~$\mathop{\Delta\xi}$} &
   \multicolumn{1}{c}{~~$\xi$} &\multicolumn{1}{c}{~~~$\mathop{\Delta\xi}$}   \\
~& 0.14&  ~&  ~&1771.494& 367.491& 429.502& 348.935& 218.154& 154.158& 213.160& 134.140& 108.841&106.506\\
~& 0.28&  ~&  ~& 240.858& 171.663&  70.223&  31.927&  42.708&  20.893&  50.913&  39.207&  51.447& 43.761\\\multicolumn{1}{c}{\multirow{1}{*}{\rotatebox{90}{Blue Galaxies}}}
~& 0.56&  ~&  ~&  40.259&  33.505&  20.414&   2.658&  17.686&   4.379&  16.873&  12.391&  19.772& 16.476\\
~& 1.12&  ~&  ~&   8.188&   3.437&   6.342&   1.239&   5.974&   2.135&   6.896&   5.297&   8.139&  3.767\\
~& 2.24&  ~&  ~&   3.336&   0.445&   2.458&   0.219&   2.162&   0.511&   2.357&   1.391&   2.791&  0.206\\
~& 4.47&   3.294&   2.496&   1.185&   0.062&   0.963&   0.036&   0.846&   0.054&   0.820&   0.167&   0.993&  0.343\\
~& 8.91&   0.461&   0.407&   0.376&   0.050&   0.323&   0.022&   0.257&   0.034&   0.244&   0.100&   0.363&  0.108\\
~&17.78&   0.369&   0.217&   0.129&   0.027&   0.104&   0.018&   0.064&   0.021&   0.061&   0.045&   0.109&  0.041\\
~&35.48&   0.122&   0.076&   0.035&   0.007&   0.029&   0.007&   0.010&   0.013&   0.011&   0.015&  ~& ~\\
\hline
~& 0.14&  ~&  ~&1203.324& 828.281& 726.580& 387.987&1000.507& 237.858& 267.370& 552.352&  ~& ~\\
~& 0.28&  ~&  ~& 741.101& 152.922& 288.323&  28.802& 400.555&  33.271& 717.431& 262.018&5641.724&467.787\\\multicolumn{1}{c}{\multirow{1}{*}{\rotatebox{90}{Red Galaxies}}}
~& 0.56& 122.181&  87.634&  99.281&  11.245&  85.996&   4.259& 125.857&   7.515& 286.505&  47.521& 643.570&213.487\\
~& 1.12&  83.521&  61.860&  20.250&   1.187&  20.914&   1.245&  30.289&   2.238&  61.773&  14.289& 193.317& 49.260\\
~& 2.24&   6.535&   3.474&   5.244&   0.301&   5.073&   0.261&   5.543&   0.408&  11.068&   2.652&  38.960&  9.931\\
~& 4.47&   5.235&   0.688&   2.000&   0.077&   1.771&   0.039&   1.717&   0.051&   2.063&   0.215&   1.801&  0.494\\
~& 8.91&   1.344&   0.156&   0.666&   0.032&   0.613&   0.023&   0.544&   0.030&   0.660&   0.111&   0.866&  0.123\\
~&17.78&   0.463&   0.060&   0.209&   0.020&   0.196&   0.017&   0.135&   0.020&   0.218&   0.048&   0.431&  0.066\\
~&35.48&   0.119&   0.033&   0.051&   0.006&   0.050&   0.007&   0.025&   0.013&   0.035&   0.015&  ~& ~\\
\hline
~& 0.14&  ~&  ~&7239.943&1692.040& 378.739& 171.435& 395.622& 151.954& 298.611& 245.189& 331.254&278.698\\
~& 0.28&  ~&  ~& 322.883& 113.649& 102.018&  22.364& 108.949&  25.905& 141.057&  71.610& 159.180&139.079\\
~& 0.56&  29.292&  11.156&  70.285&   6.320&  31.018&   2.973&  36.225&   4.991&  46.072&  27.418&  72.262& 69.431\\\multicolumn{1}{c}{\multirow{1}{*}{\rotatebox{90}{Blue-Red}}}
~& 1.12&  24.140&  12.885&  11.180&   2.338&   9.037&   1.147&  10.487&   2.172&  16.782&  14.033&  25.419& 21.754\\
~& 2.24&   3.424&   5.895&   3.949&   0.280&   3.086&   0.235&   3.076&   0.455&   4.082&   2.574&   7.629&  6.999\\
~& 4.47&   3.477&   0.747&   1.525&   0.060&   1.200&   0.034&   1.178&   0.052&   1.291&   0.189&   1.404&  0.413\\
~& 8.91&   0.701&   0.244&   0.485&   0.032&   0.388&   0.021&   0.365&   0.031&   0.395&   0.106&   0.627&  0.112\\
~&17.78&   0.371&   0.062&   0.155&   0.022&   0.102&   0.016&   0.092&   0.020&   0.120&   0.046&   0.215&  0.050\\
~&35.48&   0.100&   0.033&   0.038&   0.007&   0.004&   0.007&   0.014&   0.012&   0.021&   0.014&  ~& ~\\
\bottomrule
\end{tabular}
\textbf{Notes.} Here $r$ is the comoving distances in units of
$\mpch$; ~~$\xi$ is the two-point correlation function for a volume
limited sample; ~~$\mathop{\Delta\xi}$ is the $1\sigma$ error of
$\xi(s)$ estimated from 10 mock samples;~~The auto-correlations of
blue and red galaxies, and the cross-correlations between blue and red
galaxies, are shown in the upper, middle, and lower parts,
respectively.
\end{threeparttable}}
\end{table*} 

\begin{figure*}
\center
\includegraphics[width=0.75\textwidth,height=0.35\textwidth]{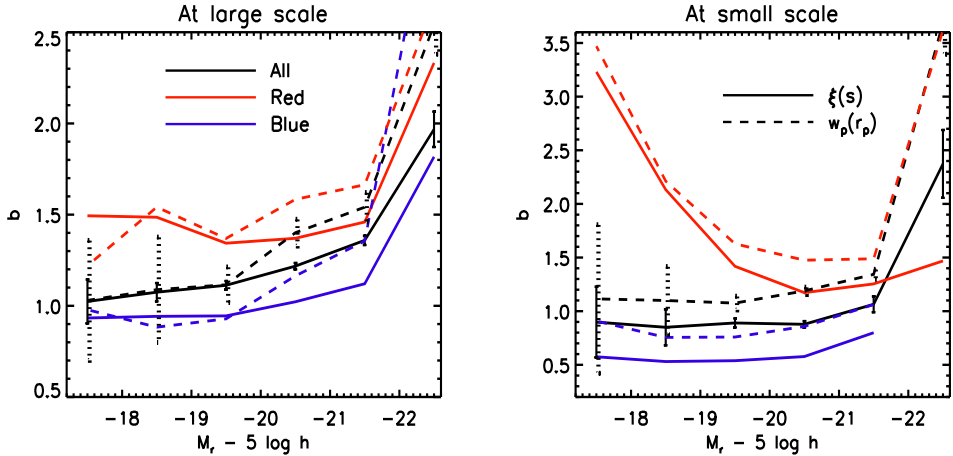}
\caption{The bias factor for SDSS DR7 galaxies as a function of galaxy
  absolute magnitude. Solid lines show the bias factors defined as the
  ratios of the measured reconstructed real-space $\xi(s)$ to that of
  dark matter over the ranges of $4\mpch < s < 20\mpch$ (left panel)
  and $0.5\mpch < s < 2\mpch$ (right panel). Dashed lines show the
  bias factors defined by the ratios of $w_p(r_p)$ between the
  redshift-space galaxies and dark matter over the ranges of $4\mpch <
  r_p < 20 \mpch$ (left panel) and $0.5\mpch <r_p < 2 \mpch$ (right
  panel).  Here the integration limit, $r_{max}$, in computing
  $w_p(r_p)$ from $\xi(r_{\rm p}, r_{\pi})$, is set to be
  $60\mpch$. Black, red and blue lines show results for all, red and
  blue galaxies, respectively. Error bars obtained from 10 mock
  samples are shown only for black solid and dashed lines.
}\label{fig:bias_dr7}
\end{figure*}

\begin{figure*}
\center
\includegraphics[width=0.75\textwidth,height=0.32\textwidth]{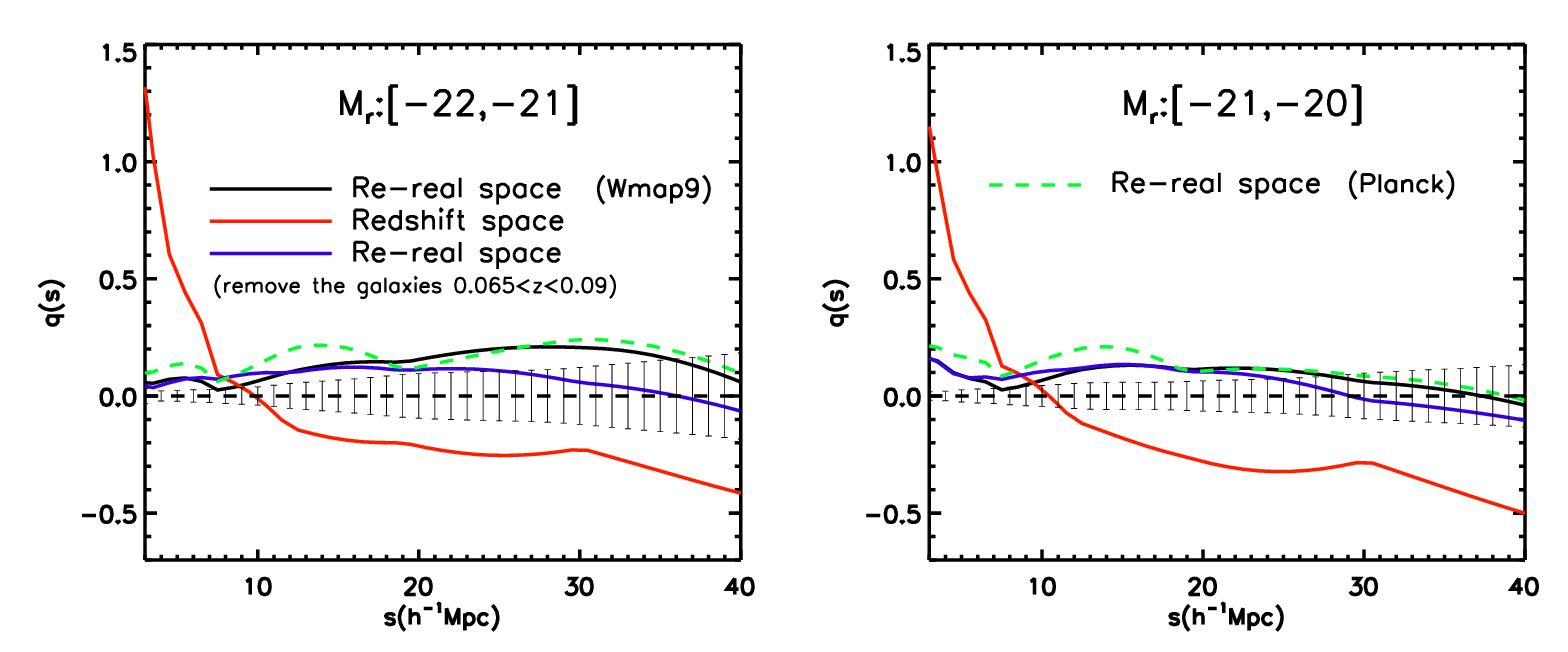}
\caption{The quadrupole-to-monopole ratio $q(s)$ for SDSS DR7 galaxy
  subsamples in the $r$-band absolute magnitude bin $M_r=$
  $\bm{[}-22.0,-21.0\bm{]}$, $\bm{[}-21.0,-20.0\bm{]}$, as
  indicated. Black and red lines correspond to the results in redshift
  and re-real spaces  (WMAP9), respectively. Blue lines are
  results in the re-real spaces in which galaxies with $0.065 \leq z
  \leq 0.09$ are removed to check the impact of the Sloan Great
  Wall. The results for the Planck cosmology are shown with green
    dashed lines.  The error bars are $1\sigma$ variances obtained
  from 10 mock re-real space samples. The dashed lines are the zero
  reference line. }\label{fig:qs_dr7}
\end{figure*}       

Fig.~\ref{fig:xiS_dr7} shows the one-dimensional 2PCFs in redshift
space (red lines) and in re-real space (black lines) for WMAP9
cosmology, for all the six magnitude samples, as indicated.
 While the green lines are results for Planck cosmology, here
  again show very good agreement with those for WMAP9 cosmology. For
comparison, the results of both the flux-limited and volume-limited
samples are shown. Note that for the two brightest samples,
flux-limited and volume limited samples are identical. For the other
samples, the correlation functions obtained from the two types of
samples are very similar, even though the samples themselves are quite
different, especially for the faint magnitude bins (see
Table~\ref{subsamp}).  Error bars for the real space correlation
function indicate the $\pm 1\sigma$ variance among the 10 re-real mock
samples described in \S\ref{sec:mock}.  All the results shown in the
figure are also listed in Table~\ref{2PCF_dr7}.  To our knowledge,
this is the first attempt to infer the real-space correlation function
of galaxies in the SDSS directly from a reconstructed real space
galaxy catalog. Note that the real space 2PCFs clearly deviate from a
simple, single power-law, revealing a clear 1-halo to 2-halo
transition on scales of $1-3 \mpch$. As demonstrated in
\S\ref{sec:mock}, this transition is more pronounced in real space
than in the projected space. It is, therefore, expected that fitting
halo occupation models directly to the real space correlation
functions presented here will provide more stringent constraints on
the galaxy-dark matter halo connection, something we will pursue in a
forthcoming paper. Finally, the lower panels of Fig.~\ref{fig:xiS_dr7}
show the ratio $\xi_1/\xi_2$, where $\xi_1(s)$ and $\xi_2(s)$ are the
two-point correlation functions in redshift space and re-real space,
respectively. This nicely shows how redshift space distortions boost
the correlation power on large scales (by about $40-50\%$ at a scale
of $10 \mpch$), while suppressing it on small scales (by about
$70-80\%$ at a scale of $0.3 \mpch$).

To study how galaxy clustering depends on galaxy color, we use the
bimodal distribution in the color-magnitude plane \citep[e.g.][]
{Str2001, Bal2004} to divide each of the luminosity samples into
``blue'' and ``red'' subsamples. Specifically, the demarcation line we
use is $(g-r) = 0.21-0.03M_r$, as is in \citet{Zehavi2011}.
Information about these subsamples is given in Table \ref{subsamp}.

Figure \ref{fig:xiS_color} shows the 2PCFs of red (red lines) and blue
galaxies (blue lines) in re-real space for different magnitude bins,
as indicated.  The result of the full sample in each magnitude bin is
also shown in each panel as black line. Green lines show the
cross-correlation functions between blue and red galaxies. The
cross-correlation is obtained by replacing $\langle DD \rangle$,
$\langle RR \rangle$ and $\langle DR \rangle$ with $\langle D_1D_2
\rangle$, $\langle R_1R_2 \rangle$ and $(\langle D_1R_2 \rangle +
\langle D_2R_1 \rangle)/2$, respectively, in Equation
\ref{eq:2pcf}. Here subscripts `1' and `2' denote red and blue
galaxies, respectively, so that $D_1 D_2$ is the number of cross pairs
between red and blue galaxies, and so on. Error bars are obtained from
the 10 mock samples.  All the data shown in this plot are also listed
in Table \ref{2PCF_color} for references. As one can see, red galaxies
exhibit higher clustering amplitude than the blue ones in the same
luminosity bin, and the cross-correlation lies in between. The
difference between red and blue galaxies appears to be larger for
fainter galaxies.

Figure \ref{fig:bias_dr7} shows the bias factors defined in the same
way as those in Figure \ref{fig:bias}. Solid lines in the left panel
show the bias factors obtained from using the values of $\xi(s)$ at
large scales, $4\mpch < s < 20\mpch$, while in the right panels, the
same type of lines show the bias factors obtained by using data on
small scales, $0.5\mpch < s < 2\mpch$.  Black, red and blue lines show
the results for all, red and blue galaxies in each $M_r$ bin,
respectively. Clearly, the bias factor depends on galaxy luminosity,
but the dependence is not in the same way for red and blue
galaxies. Overall, red galaxies have a higher bias factor than their
blue counterparts in the same luminosity bin.  The difference is the
largest for faint galaxies on small scales.  For the total and blue
populations, the bias factor on large scales increases with
luminosity. In contrast, for red galaxies, the bias factor on large
scales remains more or less constant all the way to $M_r-5\log h\sim
-21.5$, and only increases with luminosity for the brightest
galaxies. On small scales, the bias factor is quite independent of
luminosity for both the total and blue populations at $M_r-5\log h>
-20.5$, and increases with luminosity for higher luminosities.  In
contrast, the bias factor for red galaxies {\it decreases} with
increasing luminosity, especially for faint galaxies. This indicates
that faint red galaxies are preferentially satellites located in
relatively big halos, consistent with the results of \citet{Lan2016}
based on the luminosity functions of galaxies in groups.

For comparison, the dashed lines in Figure \ref{fig:bias_dr7} shows
the bias parameters obtained from the projected 2PCFs, $w_p(r_p)$,
again estimated in the same way as those for mock galaxies (see Figure
\ref{fig:bias}). The results show again that the bias parameter, $b$,
estimated from the projected 2PCF has larger errors and is biased
relative to that obtained from the reconstructed real-space $\xi(s)$,
as is demonstrated using mock samples shown in Figure \ref{fig:bias}.
This suggests that the bias parameters obtained earlier in the
literature on the basis of $w_p(r_p)$ may be significantly biased. We
will come back to a detailed analysis on this in a forthcoming paper.

Finally, we compute the quadrupole-to-monopole ratio $q(s)$ for the
SDSS DR7 galaxies.  Figure \ref{fig:qs_dr7} shows the $q(s)$ for two
luminosity samples, $M_r=[-22.0,-21.0]$ and $[-21.0,-20.0]$,
respectively. In each panel, results are shown for galaxies in both
redshift and re-real spaces using lines with different colors. as
indicated.  The error bars on top of the zero line correspond to
$1\sigma$ variances obtained from 10 mock samples in re-real space.
We see that $q(s)$ in re-real space in SDSS DR7 has a systematic
deviation from the zero line at 2-$\sigma$ level, especially for the
high-luminosity bin. This deviation may indicate that at $z\leq 0.12$
the SDSS DR7 volume still suffers from cosmic variance, likely
produced by the existence of rare large-scale structures, such as the
Sloan Great Wall. To check this we estimate $q(s)$ excluding galaxies
with redshifts $0.065 \leq z \leq 0.09$, which effectively excludes
the Sloan Great Wall.  The results are shown in Figure
\ref{fig:qs_dr7} as the blue lines.  The deviations from the zero line
are significantly reduced at large $s$.  This test result suggests
that the quadrupole-to-monopole ratio is sensitive to the presences of
large scale structures, and a much larger volume is required to get a
reliable estimate of this quantity.


  On the other hand, as discussed at the beginning of this
  subsection, the reconstruction to obtain the re-real space
  distribution of galaxies is cosmology dependent.  If the real
  universe deviates from the assumed cosmology, systematic errors can
  also be introduced in our reconstruction. The $q(s)$ for Planck
  cosmology, which are shown in Figure \ref{fig:qs_dr7} as the green
  dashed lines, do show some differences from those for the WMAP9
  cosmology.  After the removal of the Sloan Great Wall, the
deviation of $q(s)$ from zero is about $0.1$ at $s \sim 20\mpch$. This
corresponds to an under-estimate of $\beta$ by about $0.07$ in the
linear regime by WMAP9.  We will perform a detailed cosmological
  probe in a follow-up paper.


\section{SUMMARY}
\label{sec:con}

We have presented a method to correct redshift space distortions in
redshift surveys of galaxies. Adopting the method introduced in W12,
we use galaxy groups identified with the Y05 halo-based group finder
to reconstruct the large scale velocity field, which in turn is used
to correct the observed redshifts for the Kaiser effect.  The same
galaxy groups are also used to correct the Finger-Of-God (FOG) effect
produced by the virial motions of galaxies within their host dark
matter halos. Our FOG correction is based on the assumption that
satellite galaxies are an unbiased tracer of the mass profile and
velocity structure of the host halo.

To test the method, we have constructed 10 mock SDSS DR7 galaxy
catalogs, in four different spaces: redshift space (equivalent to the
observational space), Kaiser space (space in which the FOG effect is
absent), FOG space (space in which Kaiser effect if absent), and real
space (space in which redshift distortions are absent). We test the
various components of our reconstruction method by comparing the
two-point clustering statistics in these different spaces.

The contours of the two-dimensional 2PCFs $\xi(r_\rmnp,r_\pi)$
calculated in different spaces show that the clustering in our
reconstructed space is in good agreement with that in the
corresponding true space given directly by numerical simulations.  On
small transverse scales $r_\rmnp$, residual FOG effects are apparent,
which arise mainly from the uncertainties in the group finder,
including errors in group membership determinations, designations of
centrals and satellites, and halo mass assignments. We have shown,
though, that the one-dimensional 2PCF, $\xi(s)$, inferred directly
from the reconstructed real space is not significantly affected, with
deviations typically being smaller than the uncertainties arising from
cosmic variance (at least for a SDSS-like survey) for galaxies
brighter than $\rmag=-19.0$. In fact, over the range of scales $0.2
\mpch \lta r \lta 20 \mpch$, the average error on the reconstructed
real space 2PCF is less than five percent. Hence, our method is
capable of correcting redshift distortions in redshift surveys to a
level that allows for an accurate, unbiased measurement of the
real-space correlation function.

We have applied our reconstruction method to the SDSS DR7, giving 
a real space version of the main galaxy catalog which contains 396,068 
galaxies in the NGC with redshifts in the range $0.01 \leq z \leq
0.12$. This real space galaxy catalog is publicly available at
\href{url}{http://gax.shao.ac.cn/data/data1/SDSS7\_REAL.tar}.
We emphasize that the FOG correction is only statistical in nature,
and that the line-of-sight position of satellite galaxies in the
catalog have been assigned at random, in accordance with our
assumption that satellite galaxies are an unbiased tracer of the
mass distribution of their host halo. 

Using the reconstructed real space data we have shown that the Sloan
Great Wall, the largest known structure in the Universe, is not as
dominant an over-dense structure as it appears in redshift space, but
that its apparent over-density is strongly enhanced by the Kaiser
effect. We have measured the 2PCFs in reconstructed real space in
different absolute magnitude bins. They all deviate clearly from a
simple power-law, revealing a clear 1-halo to 2-halo transition. A
comparison with the corresponding 2PCFs in redshift space nicely
demonstrates how redshift space distortions boost the correlation on
large scales (by about $40-50\%$ at a scale of $10 \mpch$), while
suppressing it on small scales (by about $70-80\%$ at a scale of $0.3
\mpch$). We have also measured the real-space autocorrelation
functions of blue and red galaxies, and their across-correlations.
Using the real-space (color-dependent) $\xi(s)$, we have investigated
how the bias factor depends on galaxy luminosity and color, and how
our method provides more reliable measurements of galaxy bias factors
than the traditional method that uses the projected 2PCF, $w_p(r_p)$.

The present paper, the first paper in a series, is focused on the
methodology.  In a forthcoming paper we will use our reconstructed,
real-space SDSS galaxy catalog to study in more detail how the real
space clustering of galaxies depends on their intrinsic properties,
such as luminosity, stellar mass, color and star formation rate.  We
will also use our reconstruction method to put constraints on
cosmological parameters as well as halo occupation models.  As briefly
mentioned in \S\ref{sec:clusan}, the actual reconstruction is
cosmology dependent, as the bias parameter $b_{\rm hm}$, the halo
masses assigned to galaxy groups, and the distance-redshift relation
are all cosmology dependent. Consequently, assuming an incorrect
cosmology can result in systematic errors in our reconstruction, and
distortions in the correlation functions. We can then model such
distortions and constrain cosmological parameters by searching for the
model that gives the best reconstructed real space, so that
$\xi(r_\rmnp,r_{\pi})$ is isotropic (i.e., quadrupole-to-monopole
ratio is close to zero).


\section*{Acknowledgments}

We thank the anonymous referee for helpful comments that greatly
improved the presentation of this paper.  This work is supported by
the 973 Program (No. 2015CB857002), national science foundation of
China (grant Nos. 11128306, 11121062, 11233005, 11073017),
NCET-11-0879, the Strategic Priority Research Program ``The Emergence
of Cosmological Structures" of the Chinese Academy of Sciences, Grant
No. XDB09000000 and the Shanghai Committee of Science and Technology,
China (grant No. 12ZR1452800).  We also thank the support of a key
laboratory grant from the Office of Science and Technology, Shanghai
Municipal Government (No. 11DZ2260700).  HJM would like to acknowledge
the support of NSF AST-1517528, and FvdB is supported by the US
National Science Foundation through grant AST 1516962.

A computing facility award on the PI cluster at Shanghai Jiao Tong
University is acknowledged. This work is also supported by the High
Performance Computing Resource in the Core Facility for Advanced
Research Computing at Shanghai Astronomical Observatory. 


\end{document}